\documentclass[twocolumn,eqsecnum,aps,prd,%showpacs,preprintnumbers,
amsmath,amssymb,nofootinbib,floatfix]{revtex4}

\usepackage{bm}
\usepackage{graphicx,epsf}

\newcommand\lsim{\lesssim}

\newcommand\gsim{\gtrsim}

\newcommand\vev[1]{{\langle {#1} \rangle}}

\renewcommand\({\left(}

\renewcommand\){\right)}

\renewcommand\[{\left[}

\renewcommand\]{\right]}

\newcommand\eq[1]{Eq.~(\ref{#1})}

\newcommand\eqs[2]{Eqs.~(\ref{#1}) and (\ref{#2})}

\newcommand\eqss[3]{Eqs.~(\ref{#1}), (\ref{#2}), and (\ref{#3})}

\newcommand\eqst[2]{Eqs.~(\ref{#1})--(\ref{#2})}

\newcommand\eqreff[1]{(\ref{#1})}

\newcommand\eqsref[2]{(\ref{#1}) and (\ref{#2})}

\newcommand\eqssref[3]{(\ref{#1}), (\ref{#2}), and (\ref{#3})}

\newcommand\pa{\partial}

\newcommand\ee{\end{equation}}

\newcommand\be{\begin{equation}}

\newcommand\eea{\end{eqnarray}}

\newcommand\bea{\begin{eqnarray}}

\newcommand\mpl{M_{\rm P}}

\def\calp{{\cal P}}

\newcommand\bfx{{\mathbf x}}

\newcommand\TeV{\,\mbox{TeV}}

\newcommand\GeV{\,\mbox{GeV}}

\newcommand\MeV{\,\mbox{MeV}}

\newcommand\msun{M_\odot}

\newcommand\sub[1]{_{\rm #1}}

\newcommand\mone{^{-1}}

\newcommand\mtwo{^{-2}}

\newcommand\mhalf{^{-1/2}}

\newcommand\half{^{1/2}}

\newcommand\quarter{^{1/4}}

\newcommand\diff{d}

\newcommand\luv{{\Lambda\sub{UV}}}
\newcommand\phiend{{\phi\sub{end}}}
\newcommand{\meta}{{|\eta_0|}}

\newcommand\fnl{{f\sub{NL}}}
\newcommand\tnl{{\tau\sub{NL}}}

\begin{document}

\title{Inflation models and observation}

\author{Laila Alabidi and David H.\ Lyth}

\affiliation{Physics Department, Lancaster University,LA1 4YB}

%\date{}

    \begin{abstract}
We consider small-field models which invoke  the usual framework
 for the effective
field theory, and large-field models which go beyond that.
Present and future possibilities for discriminating
between the models are assessed, on the assumption that the primordial
curvature perturbation is generated during inflation.
With    PLANCK data, the  theoretical and observational uncertainties
on the spectral index  will be comparable, providing useful discrimination
 between small-field models. Further discrimination between models
may come later through the tensor fraction, the running
of the spectral index and non-gaussianity.
The prediction for the trispectrum in a generic multi-field inflation model
is given for the first time.
\end{abstract}

    \maketitle

        \section{Introduction}\label{intro}

        If the primordial curvature perturbation is generated
during inflation, observation constrains
the height and shape of  the         inflationary potential.
We   consider  various possibilities for
the form for the potential,
seeing to what extent present and future observation can distinguish between
them. We
cover most of the forms that have been proposed, though our citations
are far from exhaustive.

 In Section \ref{s:efolds}
we recall estimates for the number of $e$-folds  occurring after
the observable Universe leaves the horizon, which has to be specified
before a model can be constrained. In Section \ref{s:predobs} we recall
the formulas giving the spectral tilt  and the tensor fraction
generated during
 single-component inflation. We give  the present observational
constraints on these quantities as well as future projections, and we
plot the region of the $r$-$n$ plane corresponding to small-field models.
In Section \ref{effective} we recall the usual framework for effective field
theory, and in
\ref{s:smallfield} we consider small-field models based on that framework.
In Section \ref{s:naturalchaotic} we consider
Natural Inflation along with its limiting
case  $V\propto \phi^2$. We also consider the multi-field extension of that
case, recalling and extending for that purpose the multi-field formulas
for the tilt, the tensor fraction and the non-gaussianity. In
 Section \ref{s:outlook}
we see how the models might fare after PLANCK data is
available and we conclude in Section \ref{conclusion}.

        \section{The number  of $e$-folds}

\label{s:efolds}

Observation provides a direct constraint on inflation  only after
the observable Universe leaves the horizon. We may call inflation
during this era `observable inflation' to distinguish it from the
possibly very large amount of inflation occurring earlier. In
order to make predictions, we need the number $N(k)$ of
$e$-folds of slow-roll inflation,  remaining after a scale $k$
leaves the horizon, where
 $k$ is the coordinate wavenumber corresponding to physical wavenumber
$k/a$.\footnote
{Standard results about early-universe cosmology and observation
 are described for instance in  \cite{treview,book}.}
 The biggest scale of interest is roughly $k=a_0H_0$ (where $H\equiv
\dot a/a$ is the Hubble parameter
and the subscript denotes the present) and we denote
$N(a_0H_0)$ by simply $N$. While cosmological scales leave the horizon we
 assume almost-exponential inflation giving
\be
N(k) = N - \ln(k/a_0H_0)
\label{nofk05}
.
\ee

  To determine $N$ one needs to know something about the
  history of the Universe between the end of
  slow-roll inflation and the onset of nucleosynthesis.
 To be precise, assuming Einstein gravity, one needs the pressure $P$ as a
  function of the energy density $\rho$. There is also a weak dependence on
  the value of the
  Hubble parameter during inflation, equivalent with Einstein gravity to the
  height of the inflationary potential.

  Assume first radiation domination ($P=\rho/3$) from the end of  inflation
  to the onset of nucleosynthesis. Then
        \bea
             N
            &=& 61  -  \ln\(\frac{10^{16}\GeV}{V^{1/4}}\)
            \label{nefolds}
            \,.
        \eea
Here $V$
is the inflationary potential,  which for the present purpose can be taken
to be constant.

        {}Scales of cosmological interest correspond to
        \be
            0 \lsim  \ln\frac{k}{a_0H_0} \lsim 14
            \,,
        \ee
where the upper limit is the scale enclosing matter with mass $10^6\msun$.
They have to leave the horizon during inflation which gives the lower bound
$            N\gsim 14$.
Observation   requires $V\quarter\lsim 10^{16}\GeV$ and the onset
        of nucleosynthesis at temperature $T\simeq 1\MeV$
        requires  $V\quarter\gsim 1\MeV$, which places  the last
        term of \eq{nefolds}
in the range $0$ to $-40$.

Now we consider other possibilities for $P(\rho)$.
        {}The biggest reasonable pressure \cite{andrewn}
is $P=\rho$, corresponding to domination by
        the kinetic term of a scalar field (kination).
If kination dominates from the end of inflation to the onset of
nucleosynthesis,  the above estimate of $N$ is increased
by $\frac13 \ln(V\quarter/1\MeV)<14$.
If instead there is matter domination ($P=0$)
during that era, the estimate is decreased by the same amount.
%We conclude
%that if any combination of kination, matter and radiation dominates between
%the end of inflation and the onset of nucleosynthesis,
%\be
%47 < N < 75
%\label{Nest1}
%\,.
%\ee

Finally, we consider the possibility that significant inflation occurs
after slow-roll ends.
The most plausible way of arranging this is to invoke one or more bouts
of Thermal Inflation \cite{thermal},
taking place at a relatively low energy scale. Each bout would reduce the
estimate $N$ by $10$ or so.

{}From this discussion,   we learn that in principle,
$N$ might in principle be anywhere in the range
\be
14< N < 75
\label{Nest2}
\,.
\ee
On the other hand, most early Universe scenarios give $0< P < \rho/3$,
 and most of the models of inflation that we shall discuss give
 $V$  near the top of its allowed range. In that case,
taking $V$ at the very top,  we have
 $47< N <62$ corresponding to
\be
N= 54 \pm 7
\label{Nest3}
\,,
\ee
and a fractional uncertainty
\be
\frac{\Delta N}{N} = 0.13
\label{delNoverN}
\,.
\ee
Taking  instead  $V\quarter\sim 10^{10}\GeV$ (which corresponds to
$H\sim \TeV$ and is the lowest value usually considered)
we have $N=34\pm 7$ and a fractional error $\Delta N/N=0.20$.

This discussion leads to a very important conclusion.
 {\em For any reasonable inflation scale, and post-inflation
pressure in the standard range $0< P < \rho/3$, the fractional error
in $N$ is of order 10 to $20\%$.} As we shall see, the corresponding
uncertainties in the predictions are of the same order in a wide range of
models.

        \section{Prediction and observation} \label{s:predobs}

\subsection{Slow-roll inflation}

        We assume Einstein
gravity, and take the fields to be canonically normalized.
Until Section \ref{s:multi} take the slow-rolling inflaton to have
a single component.
  During inflation,  the potential $V(\phi)$ depends only on the
inflaton field $\phi$. It is supposed that the  field equation
\be
\ddot\phi + 3H\dot\phi + V' = 0
\ee
is  well-approximated by
\be
3H \dot \phi = -V'
\label{phidot}
\,,
\ee
and that the energy density $3\mpl^2H^2=V+\frac12\dot\phi^2$ is
slowly varying on the Hubble timescale. (We are defining $\mpl=(8\pi G)\mhalf
=2.4\times 10^{18}\GeV$.)
These conditions imply
\be
3\mpl^2 H^2 \simeq V
\label{vofh}
\,.
\ee
and the flatness
conditions
        \be
            \epsilon\ll 1 \qquad  |\eta|\ll 1 \label{flat}
\,,
        \ee
   where
\bea
\epsilon &\equiv&  \frac12\mpl^2(V'/V)^2 \label{epsilondef} \\
\eta &\equiv&   \mpl^2V''/V \label{etadef}
\eea
 Conversely, slow-roll
inflation can usually take place on any portion of the potential
satisfying the flatness conditions. In the slow-roll approximation
$\epsilon $ is slowly varying; \be H\mone \dot \epsilon =
2\epsilon(2\epsilon-\eta) \label{epsdot} \,. \ee For future
reference we note that $\dot\epsilon$ is positive if $\ln V$ is
concave-downward and negative if $\ln V$ is concave-upward.

To obtain the predictions, one needs the field value $\phi(k)$ when
a given scale leaves the horizon. {}From \eqs{vofh}{phidot}
it  is related to the number of $e$-folds by $dN(k)/d\phi=\mpl^{-2}V/V'$
and we focus on the biggest scale $k=a_0H_0$. When this scale leaves the
horizon, the field value $\phi_*$ is given by
        \be
            N = \mpl^{-2} \int^{\phi_*}_{\phi\sub{end}}
\(
\frac{V}{V'}
\)\diff \phi =\mpl\mone  \int^{\phi_*}_{\phi\sub{end}}
\frac{\diff\phi}
{\sqrt{2\epsilon(\phi)}}
            \,,
            \label{nofk}
        \ee
        {}where $\phi\sub{end}$ is the value at the end of slow-roll
        inflation and a star denotes horizon exit for the biggest  scale.
If $\ln V$ is concave-downward, $\epsilon$ increases with time.
Then the value of $N$ will typically be insensitive to
$\phi\sub{end}$, making the model more predictive. The only
exception to this rule that arises in practice is the potential
 $V=V_0-\frac12m^2\phi^2$ with $V_0$ dominating.

In Figure \ref{rlog} we  plot in the $\log r$-$n$ plane
  the line
$\eta=2\epsilon$, to the left of which $\log V$ is
concave-downward,
 and the line $\eta=0$, to the left of which $V$ itself
 is concave-downward.
For $r\lsim 10\mtwo$ these lines
practically coincide. In Figure \ref{rlinear} we
 plot the same  lines in the $r$-$n$
plane.

In the vacuum, $V=0$.
We shall consider both non-hybrid models, where the inflationary value of
$V$ is generated almost entirely by the displacement of the inflaton
field from its vacuum, and hybrid models where it is generated almost
entirely by the displacement of some other (waterfall) field.
In non-hybrid models, $\epsilon$ increases with time and
inflation ends when one of the flatness conditions fails, after which $\phi$
goes to its vacuum expectation value (vev). In some
 hybrid models, $\epsilon$ decreases with time
($\log V$ concave-upward), and inflation ends only when the
waterfall field is destabilized. In other hybrid inflation models,
 $\epsilon$ increases with time ($\log V$ concave-downward), and {\em slow-roll}
inflation may end before the waterfall field is destabilized through the
failure of  one of the flatness conditions.
If that happens,  a few  more $e$-folds of inflation can take place
while the inflaton  oscillates about its vev (locked inflation \cite{locked}),
until the amplitude of the oscillation becomes low enough to destabilize the
waterfall field.

\subsection{The curvature perturbation}

The vacuum
fluctuation of the inflaton generates a
practically gaussian perturbation, with spectrum
$\calp_\phi(k) = (H_k/2\pi)^2$ where the subscript $k$ indicates horizon
exit $k=aH$. This perturbation  generates at horizon exit
 a practically gaussian time-independent contribution to the
curvature perturbation
with  spectrum \cite{specpred}
\be
\calp_\zeta = \frac1{24\pi^2\mpl^4} \frac V \epsilon
\label{spec}
\,.
\ee
Subsequently, the perturbations of additional light fields may generate
an additional contribution to the curvature perturbation,  which come to
dominate so that the inflaton contribution is irrelevant. In that case
the only constraint on the form of the slow-roll inflaton potential
is that \eq{spec} is {\em below} the observed value. Here we suppose that
instead \eq{spec} gives the dominant value.

Assuming that $\eta$ is slowly varying, the spectral tilt $n-1\equiv d\ln
\calp_\zeta/d\ln k$ is given by \cite{ll92}
\be
n-1 = 2\eta -6\epsilon
\label{ninf0}
\,.
\ee
(We will refer to the tilt $n-1$ instead of to the spectral index $n$
since  more direct  physical significance.)

In \eqs{spec}{ninf0}  the potential and its derivatives are to be evaluated at
horizon exit for the scale $k$ under consideration. We shall
evaluate them
at the single scale
$k=a_0H_0$, corresponding to $\phi_*$ given by \eq{nofk}.
For the forms of the potential that we consider, this is good enough
at least with present data.
 (As we see later future observation might detect the  scale
dependence of $n$ (running). We are not considering the
  running mass model, which gives strong running of $n$ that is already
  constrained by present observations  \cite{cl,ourrunning}.)

 The spectrum is measured with good accuracy
as $\calp_\zeta= (4.9\times 10^{-5})^2$, which with the prediction \eq{spec} requires
        \be
 \frac{V^{1/4}}{\epsilon\quarter} = 0.027 \mpl = 6.6\times10^{16}\GeV
            \,.
\label{cmbnorm}
        \ee
At present, measurements of the tilt are consistent with
zero. Taking the tensor fraction $r$ to be negligible,
which is the prediction of a wide class of models,
a         fit  using WMAP CMB data and the SDSS galaxy survey
        \cite{sdsswmap}  gives
            \be
              %  n=    0.977 + 0.039 - 0.025
n= 0.980 \pm  0.020
\label{b1}
                \,.
            \ee
       {}A fit using instead WMAP CMB data and the 2dFGRS galaxy survey gives
        \cite{2dfwmap}
        \be
            n = 0.956 \pm 0.020
            \,.
        \ee
A fit using WMAP and BOOMERANG CMB data and the SDSS and 2dFGRS galaxy surveys
gives \cite{wbs2}
\be
n= 0.950 \pm 0.020
\,.
\label{b3}
\ee
These bounds are all  at 68\% confidence level, and all three of them
are compatible at this level.
        For the purpose of illustrating our method
we use the results of the first group, for $n$ and also for the
tensor fraction that we come to next.

\subsection{The tensor fraction}

The  prediction for the tensor fraction is\footnote
{We are adopting the currently-favoured definition using the
 spectra. An earlier definition using  the CMB quadrupole
corresponded to $r=12.4\epsilon$.}
        \be
            r = \( \frac{V^{1/4}}{3.3\times 10^{16}\GeV} \)^4
            \label{rofv}
            \,,
        \ee
which with the cmb normalization \eqreff{cmbnorm} is equivalent to
\be
r=16\epsilon
\label{rofepsilon}
\,.
\ee
Another  expression for $r$ involves the field variation. Scales on
which the  tensor can be observed leave the horizon during about 4 $e$-folds,
starting  with the exit of the scale
$k=a_0H_0$. During such a brief era $\epsilon$
will be practically constant. {}From \eqs{rofv}{nofk}, it follows that the
variation $\Delta \phi_4$  during the four $e$-folds is
 related to $r$ by \cite{mygrav}
        \be
            r = \( \frac {\Delta\phi_4}{\sqrt{2}\mpl} \)^2
            \label{rbound1}
            \,.
        \ee

If  $\epsilon$ is  continuously increasing with time ($\log V$
concave-downward), \eqss{rbound1}{rofv}{nofk} give the  strong
bound \cite{bl},
        \be
            r < \frac8{N^2}
\( \frac {\Delta\phi_N}{\mpl} \)^2
= 0.0032  \( \frac{50}{N} \)^2 \( \frac {\Delta\phi_N}{\mpl} \)^2
\label{rbound2}
            \,,
        \ee
where $\Delta\phi_N$ is the total field variation.
 As we remarked earlier, the condition that $\epsilon$ be continuously
increasing usually ensures also that $N$ is
practically independent of $\phi\sub{end}$.
We shall refer to models with
$\Delta\phi_N<  \mpl$ as small-field models, and the rest as large-field
models. 

\begin{figure}
        \centering
        \includegraphics[angle=270,width=0.9\columnwidth,totalheight=2.5in]{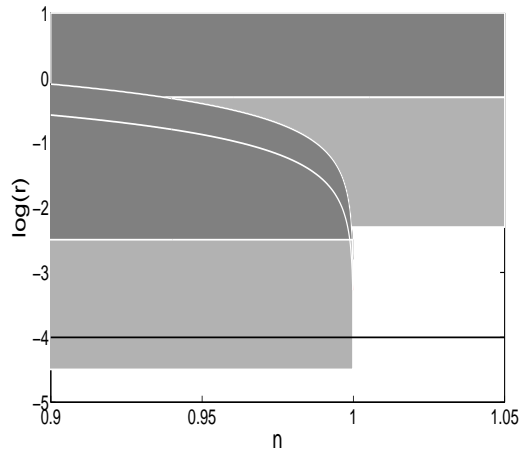}
            \caption{The large-field region corresponding to
            $\Delta\phi_N >\mpl$
            is shaded in dark grey. The region corresponding instead to  $\Delta\phi_N>0.1\mpl$
            is lightly shaded. To be eventually observable $r$ should be above
the  black horizontal line.
%barbed line corresponds to the %Planck, Clover and lower eventual
%            limits on the detectability of the tensor to scalar ratio
}
        \label{rlog}
        \end{figure}

In Figure  \ref{rlog}, the two curved lines
 divide the $\log r$-$n$ plane into three regions, according to whether $V$
and $\log V$ are concave-upward or concave-downward while cosmological scales
leave the horizon. To  the right of the
rightmost line  $\ln V$ is concave-upward while to the left of the leftmost 
line $V$ is concave-downward. If the concave-upward -downward behaviour
persists till the end of slow-roll inflation, the right-hand region is
inhabited exclusively by hybrid
 inflation models,  since otherwise inflation would never end.
%, and to be under good
%theoretical control they need $\Delta\phi_N\ll \mpl$. We conclude that
%{\em hybrid
%inflation is unlikely to give an observable  tensor fraction}.

Still assuming that the concave-upward or -downward characterisation persists
after cosmological scales leave the horizon, 
 Figure \ref{rlog} shows  (taking $N=50$) the 
 small-field region defined by $\Delta\phi_N<\mpl$ as well as a reduced
 small-field region defined by $\Delta\phi_N< 0.1\mpl$.
  (In the right-hand area where \eq{rbound1}
applies, we have  assumed that $\Delta\phi_4\sim \Delta\phi_N$,
otherwise the small-field region shrinks.)
As we shall see, for inflation models invoking the usual framework of effective
field theory, the small-field condition
 $\Delta\phi_N<\mpl$ usually  has to be
satisfied by some orders of magnitude. The only
exception is if the inflaton is a modulus field, but even there
$\Delta\phi_N/\mpl$ is usually well below 1.
In Figure \ref{rlog} we
 show how the `small-field' region shrinks if it is
defined instead by  $\Delta\phi_N<0.1\mpl$.\footnote
{The small- large-field terminology was originally introduced in
 \cite{dkk} using the older definition of the Planck scale
$m^2\sub{P}=8\pi\mpl^2$. Defining the small-field regime by
$\Delta\phi_N<m\sub P$ it would expand up to $r=0.08$,  but as we have just
noted effective field theory  models in practice satisfy $\Delta\phi_N<\mpl$
rather well.}
%The older definition is also inappropriate in that it introduces $8\pi$
%factors into crucial relations like \eqssref{hofv,flat,nefolds}.}

\begin{figure}
        \centering\includegraphics[angle=270,width=1.0\columnwidth,totalheight=2.5in]{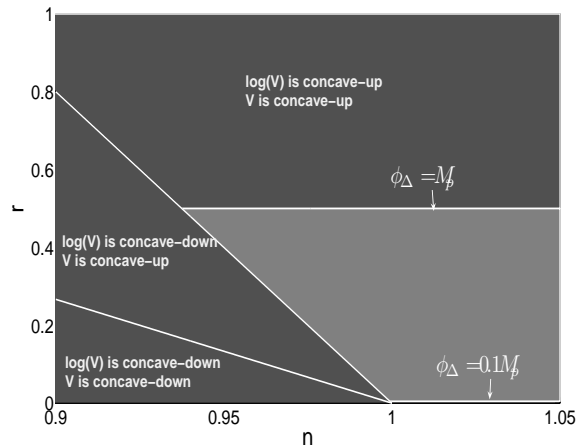}
        \caption{Figure \ref{rlog} is plotted on a linear scale.
%        The  region  corresponding to $\Delta\phi_N<0.1\mpl$ is entirely
%        invisible on this plot, and so is the left-hand part of the region
%        $\Delta\phi_N<\mpl$ which we are referring to as small-field.
}
    \label{rlinear}
 \end{figure}

In Figure \ref{rlinear} we repeat the plot of Figure \ref{rlog} using
a linear scale for $r$.
This is the plot usually seen in the literature,
but  it loses important information because a
region of very small $r$ is invisible. Following \cite{dkk} the
three regions are  usually labeled, from left to right,
`small-field', `large-field' and `hybrid', but the first labeling
is inappropriate since
the bound  \eqreff{rbound2} makes the small-field part of the left-hand region
invisible in the linear plot. Also,
hybrid models are not confined to the third region.\footnote
{Some time after the first version of the
present paper was released, \cite{kr2} appeared. It gives both the
$r$-$n$ plot and the $\log r$-$n$ plot, but the bounds
\eqsref{rbound1}{rbound2} are  not noted.}

\begin{figure}
        \centering
        \includegraphics[angle=270,width=1.0\columnwidth]{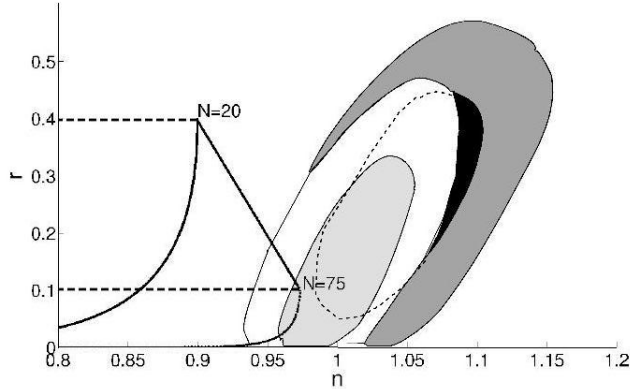}
        \caption{The curved lines are the Natural Inflation predictions for $N=20$
        and $N=75$, and the horizontal lines are the corresponding multi-component
        Chaotic Inflation predictions. The junction of each pair of lines corresponds
        to single-component Chaotic Inflation.  The regions allowed by observation
        with various assumptions are taken from \cite{tegmarketal}.}
                                    \label{rvsn}
                                \end{figure}

Now we come to observational constraints involving $r$.
If $r$ floats  in a fit to data,  $n$ should also float
 since no known inflation model gives negligible tilt with a significant
tensor fraction.  At present observation
gives
        something like $r<0.5$, $95\%$ confidence level. {}From \eq{rofv}
this requires
\be
V\quarter \lsim 3\times 10^{16}\GeV
\label{vmax}
\,.
\ee
One can also  plot the allowed region
        in the $r$-$n$ plane. This is shown in Figure \ref{rvsn} for the
data  analysis that we are using.

Absent a detection, the
 bound on $r$ will come down dramatically in the future.
Data from PLANCK \cite{planck} will give $r\lsim 0.05$.
        Clover \cite{clover}
will give something like $r<10^{-2}$ and the eventual
        limit may  be \cite{rlimit}
something like $r<10^{-4}$. Values of $r$ as low is this have not
previously been contemplated in connection with constraints on
models of inflation. In the following we will be showing some
model predictions in the $\log r$-$n$ plane.

The condition $r>10^{-4}$ for the tilt to be eventually observable is
equivalent to $V\quarter>3\times 10^{-15}$. As far as the formula
\eqreff{nefolds} is concerned, this is practically at the top of the allowed
range. We conclude that {\em for models giving an observable tensor fraction,
the estimate \eqreff{Nest3} of $N$ will  hold if
$0<P<\rho/3$ between the end of  inflation and nucleosynthesis.}

Finally, we note that for
small-field models, the tilt if it is  big enough to observe will be
accurately given by
\be
n-1 = 2\eta
\label{nofeta2}
\,.
\ee
This means that {\em for small-field models, the tilt  measures the second
derivative of the potential}. A concave-upward potential corresponds to
positive tilt and a concave-downward potential  to  negative tilt.

        \section{Effective field theory}

\label{effective}

In the usual applications of field theory
 the relevant scalar fields are
close to the  fixed point of the relevant symmetries.
 By `close' we mean that the distance to the origin in field space
is  much less than the ultra-violet cutoff $\luv$ of the effective
field theory under consideration. Starting with some fields and a
symmetry group, the  action is usually supposed to contain all
possible terms not forbidden by the symmetry, including
non-renormalizable terms. The coefficients of non-renormalizable
terms have by definition negative energy dimension. These
coefficients  parameterize physics beyond the cutoff $\luv$, and
as discussed for instance in \cite{weinberg} they
 are usually supposed to be of order 1 in  units of the cutoff.

The  cutoff presumably  cannot be bigger than $\mpl$. If it is smaller,
 the effective field theory
might be obtained by integrating out some heavy fields in
an effective theory with a bigger cutoff. Then   a definite
contribution to the coefficients of the non-renormalizable terms can be
calculated, which hopefully will be dominant.  If instead $\luv\sim \mpl$,
there is presently no good way of estimating these coefficients.

The non-renormalizable terms  are negligible
in  most of the usual applications of
field theory, but in some situations one or more
low-order non-renormalizable terms may be significant.
If the  presence of some non-renormalizable  term
is unwelcome, a symmetry might be invoked to forbid or suppress it.
Alternatively an argument from string theory or some setup with large
extra dimensions might be invoked.

This is the usual setup for effective field theory, normally employed
when considering what future colliders and detectors might find
as well as in many discussions of the early Universe.
Of course  one is  free  instead to  write down a lagrangian
which does the job in hand without
providing any justification for terms which have been omitted
(apart from the requirement that they are not effectively generated by
quantum effects).
That is the view usually taken in  for instance discussions of alternative
gravity theories involving one or more  scalar fields.

Both points of view have been taken in connection with
inflation model-building; the usual setup for effective field theory, and
the view that the required lagrangian need not be justified.
When considering  small-field inflation models one usually takes
  the former,
as reviewed for instance in \cite{treview,dl,bl}.
Its virtue, here as in other contexts, is that the possibilities for
model-building are relatively constrained so that one can exhibit some
fairly well-defined examples.

Whichever view one takes,  the energy density should be below the
cutoff;
\be
V\quarter < \Lambda\sub{UV}
\label{vlambdabound}
\,.
\ee
In most of the models that we encounter, $V\quarter$ is not
many orders of magnitude below its  maximum value $3\times 10^{16}\GeV$
which requires that $\Lambda\sub{UV}$
 is not too far below that value.
In the context of Einstein gravity $\Lambda\sub{UV}$ cannot exceed
$\mpl$. Following the usual practice we focus on the maximum
value $\luv\sim \mpl$ which turns
out to be advantageous for inflation
model-building.

In the effective field theory approach, the
 tree-level potential
is expanded as a power series in the fields
keeping only a few low-order terms.
 If the
 symmetries relevant for the inflaton are not spontaneously
broken, the fixed point lies on the trajectory and $V'$ vanishes there.
In any case, assuming that $V'$ vanishes at some point and choosing that as
the origin,
the power series expansion for the tree-level potential is
\be
V = V_0 \pm \frac12 m^2\phi^2 + M\phi^3 +
\frac14\lambda\phi^4 + \sum_{d=5}^\infty
\lambda_d \Lambda\sub{UV}^{4-d} \phi^d
\label{vseries}
\,,
\ee
with a symmetry often forbidding all odd terms.
The coefficients can have either sign but $\lambda$ and $\lambda_d$ are usually
taken to be positive and we are adopting the convention that $m^2$ is
positive.  If the non-renormalizable couplings $\lambda_d$ are
 of order 1, we need
\be
\phi\ll\Lambda\sub{UV}
\label{phicond}
\,,
\ee
 for   this series to
be under control. Then, barring a cancellation between terms,  $V_0$
has to  dominate and the flatness conditions
limit the magnitude  of each term.

For the  non-renormalizable terms,
 $|\eta|\ll 1$ with \eqs{rofv}{rbound1} requires \cite{treview}
\be
|\lambda_d| \ll 10^{-8} \(\frac {\Lambda\sub{UV}}{\mpl} \)^{d-4}
\( \frac{3\times 10^{16}\GeV}{V\quarter} \)^{2(d-4)}
\label{lambdacon}
\,.
\ee
For a given $V$ this constraint is loosest with our adopted value
$\luv\sim \mpl$.\footnote
{This constraint is stronger than \eq{vmax} if $\luv/\mpl > 
(3\times 10^{16}\GeV
/\mpl)^{(2d-4)/(d-4)}$.} It is then  compatible with  $|\lambda_d|\sim 1$
for all values of $d$
  for a low inflation scale  $V\quarter<3\times 10^{10}\GeV$.
More generally it is incompatible with  $|\lambda_d|\sim 1$ only
for the first few values, provided that $V\quarter$ is significantly below
its  maximum possible value. To forbid the first few values one can invoke
a discrete symmetry.

For the  quadratic and quartic terms, $|\eta|\ll 1$ requires  \cite{treview}
(using in the second case also \eqs{rofv}{rbound1})
\bea
m^2\mpl^2&\ll& V \\
|\lambda| &\lsim&  10^{-14}
\,.
\eea
The second  constraint is satisfied
 in a supersymmetric theory
if $\phi$ is  taken to be what is called a flat direction.
(Flat directions can  also be present in non-supersymmetric conformal
field theory \cite{paul}.)
 Regarding the
first constraint,  if the mass in a globally supersymmetric theory
is set to zero,  supergravity generates \cite{mass} a value\footnote
{If there are large extra dimensions this assumes that the
inflaton and the source of supersymmetry breaking live on different branes.
Otherwise \cite{myextra} $m^2$  is bigger by a factor
$(\luv/\mpl)^2$.}
\be
 m^2\sim  V/\mpl^2
\label{msugra}
\,,
\ee
giving a contribution $|\eta|\sim 1$.
That this  marginally violates the flatness condition
 has been called the $\eta$ problem. It may be
solved either by modest fine-tuning or by going to a non-generic supergravity
theory. More generally, it is possible
\cite{ewanmulti,pseudonat}   to control
 the mass by making the inflaton a pseudo-Nambu-Goldstone boson.

The condition on the mass becomes   problematic if one supposes that the
 inflaton is  a fermion condensate in a generic non-supersymmetric
field theory.
%, which is entertained  for instance in \cite{hector}.
%For the theory to make sense one needs $V\lsim \Lambda\sub{UV}^4$, but the
% strong interaction needed to form the condensate will presumably
%lead to $|\lambda_d|\sim 1$ leading to a  violation of \eq{lambdacon}.
In such a theory one expects $m$ roughly of order $\luv$, whereas the condition
 on the mass together with  $V<\Lambda\sub{UV}^4$ gives
\be
\frac{m^2}{\Lambda\sub {UV}^2}\ll \frac V{\mpl^2\Lambda\sub {UV}^2} \lsim
10^{-4}\( \frac{V\quarter}{3\times 10^{16}\GeV} \)^2
\,.
\ee

This discussion, based on \eq{vseries} with the expectation
$|\lambda_d|\sim 1$, is appropriate for typical fields.
According to
string theory there are likely to be special fields (moduli) whose potential is
of the form
\be
V(\phi) = V_0 f(\phi/\mpl)
\label{vmod}
\,,
\ee
with $f$ and its low derivatives roughly of order 1 at a typical point in the
range $0<\phi<\mpl$. Expanded about a given point, this gives
non-renormalizable terms with coefficients of order $\pm V_0/\mpl^4$.
Taking the point to be a maximum or a minimum so that the expansion has
the form \eqreff{vseries}
\be
m^2\sim V_0/\mpl^2
\label{msquaredofvzero}
\,.
\ee
In the simplest setup, $m\sim \TeV$ corresponding to $V_0\quarter\sim
10^{10}\GeV$ but that is not supposed to be mandatory.

There  is usually supposed to be more than
one modulus. If the effective field theory is supersymmetric,
the moduli come in pairs (usually taken as components of a complex field)
of which one, the `axion'  is a PNGB  whose potential is periodic with
(from \eq{vmod}) a  period of order $\mpl$.

\section{Small-field models}

\label{s:smallfield}

In this section we deal with models of inflation based on the effective
field theory approach of the last section. To keep the non-renormalizable
terms under control we limit  the inflaton field to
the range $0<\phi<\mpl$, making the models
 automatically small-field models.

\subsection{New and modular inflation}
\label{newmod}

\begin{figure*}
   \begin{minipage}[t]{0.5\linewidth}
 \centering\includegraphics[angle=270, width=3in]{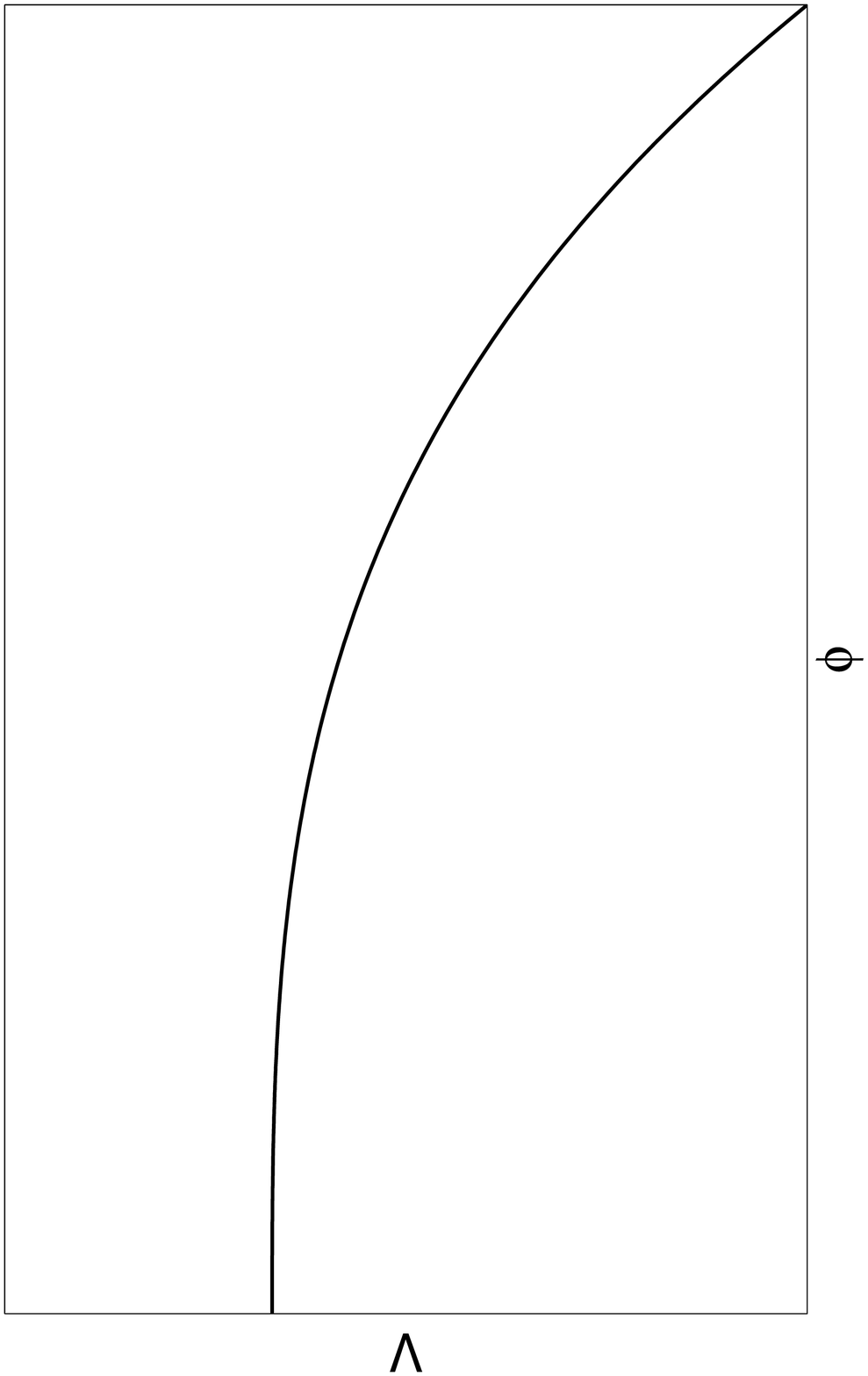}
  \caption{New inflation, modular inflation, etc.}
    \label{smallv1}
\end{minipage}%
   \begin{minipage}[t]{0.5\linewidth}
\centering\includegraphics[angle=270, width=3in]{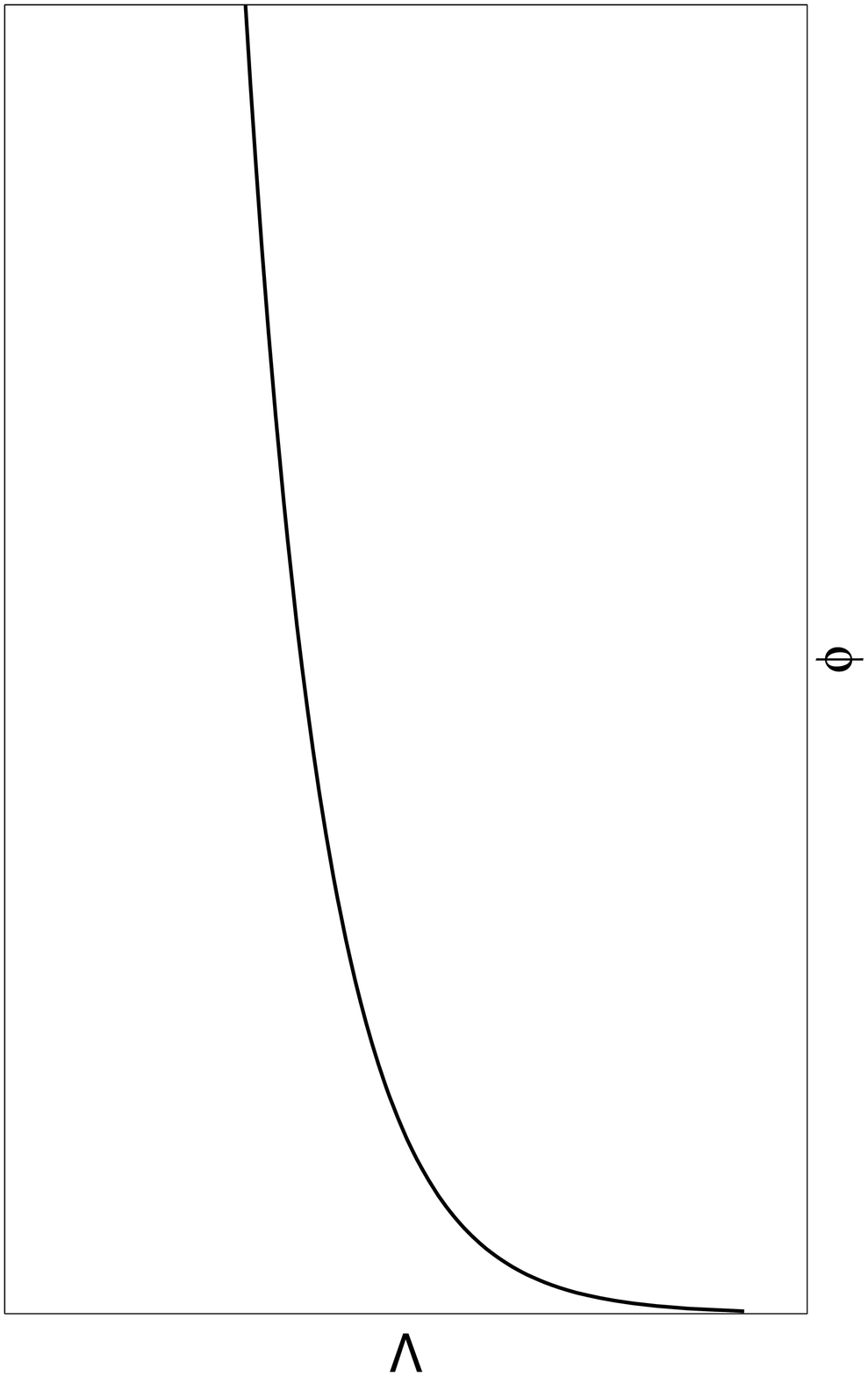}
\caption{$F$- and $D$-term inflation, mutated hybrid,
                        $D$-branes etc.}
                                    \label{smallv2}
                            \end{minipage}\\[20pt]
                            \begin{minipage}[t]{0.5\textwidth}
 \centering\includegraphics[angle=270,width=3in]{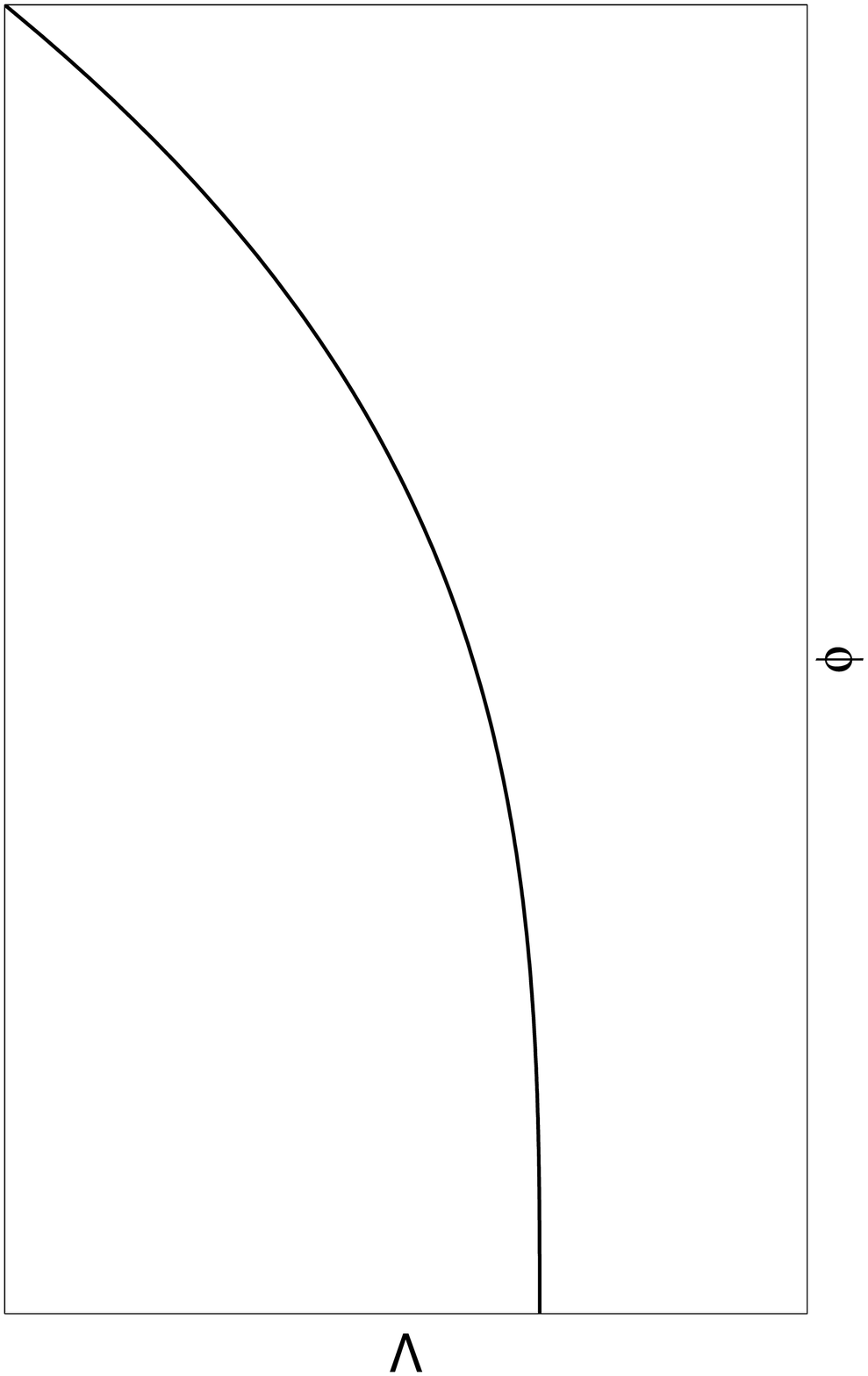}
                                \caption{Tree-level hybrid etc.}
                                    \label{smallv3}
                            \end{minipage}%
                            \begin{minipage}[t]{0.5\textwidth}
              \centering\includegraphics[angle=270,width=3in]{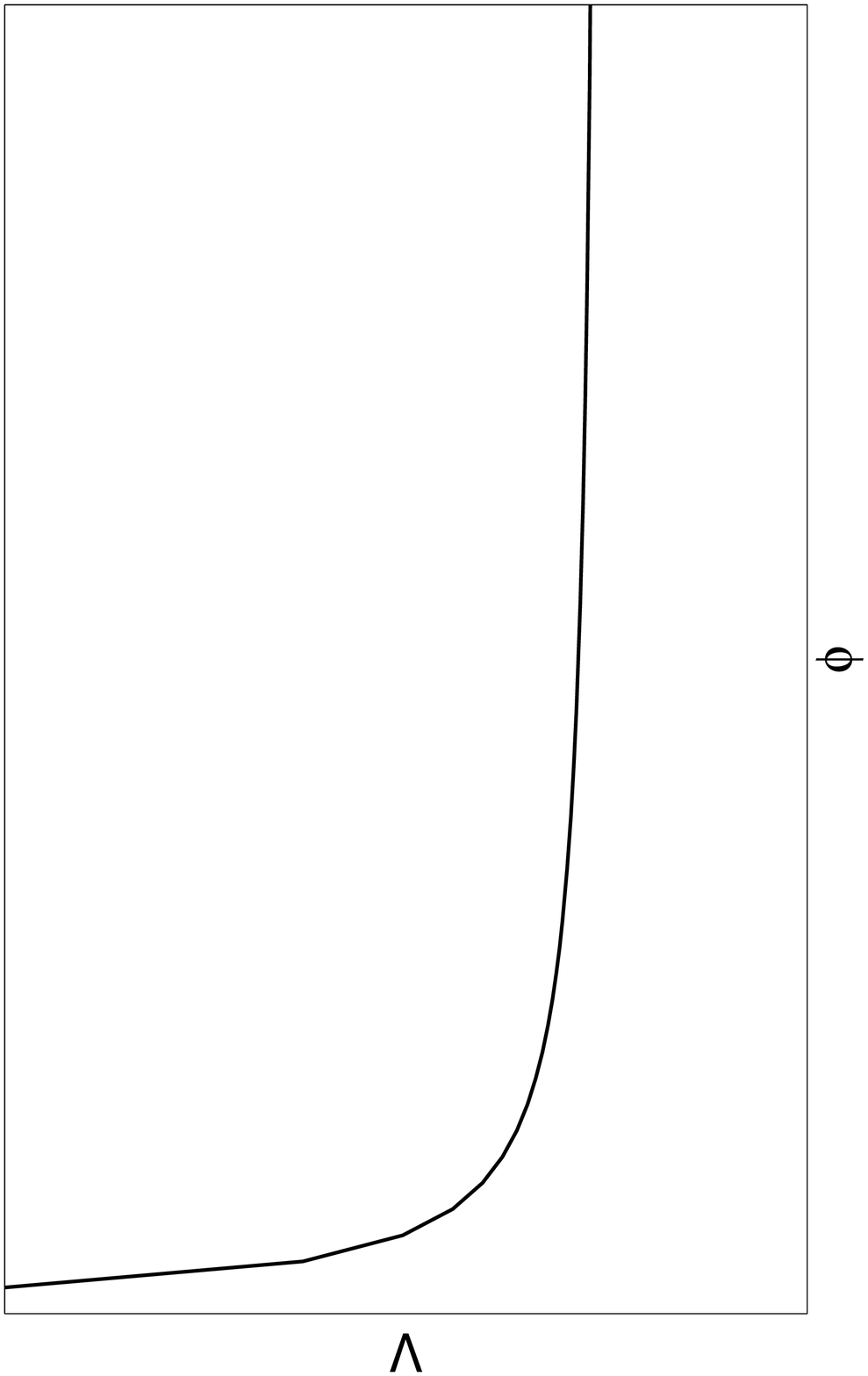}
           \caption{Dynamical supersymmetry breaking, $D$-branes etc.}
\label{dbranes}
                            \end{minipage}\\[20pt]
                            \begin{center}
                            \begin{minipage}[t]{0.5\textwidth}
        \centering\includegraphics[angle=270, width=3in]{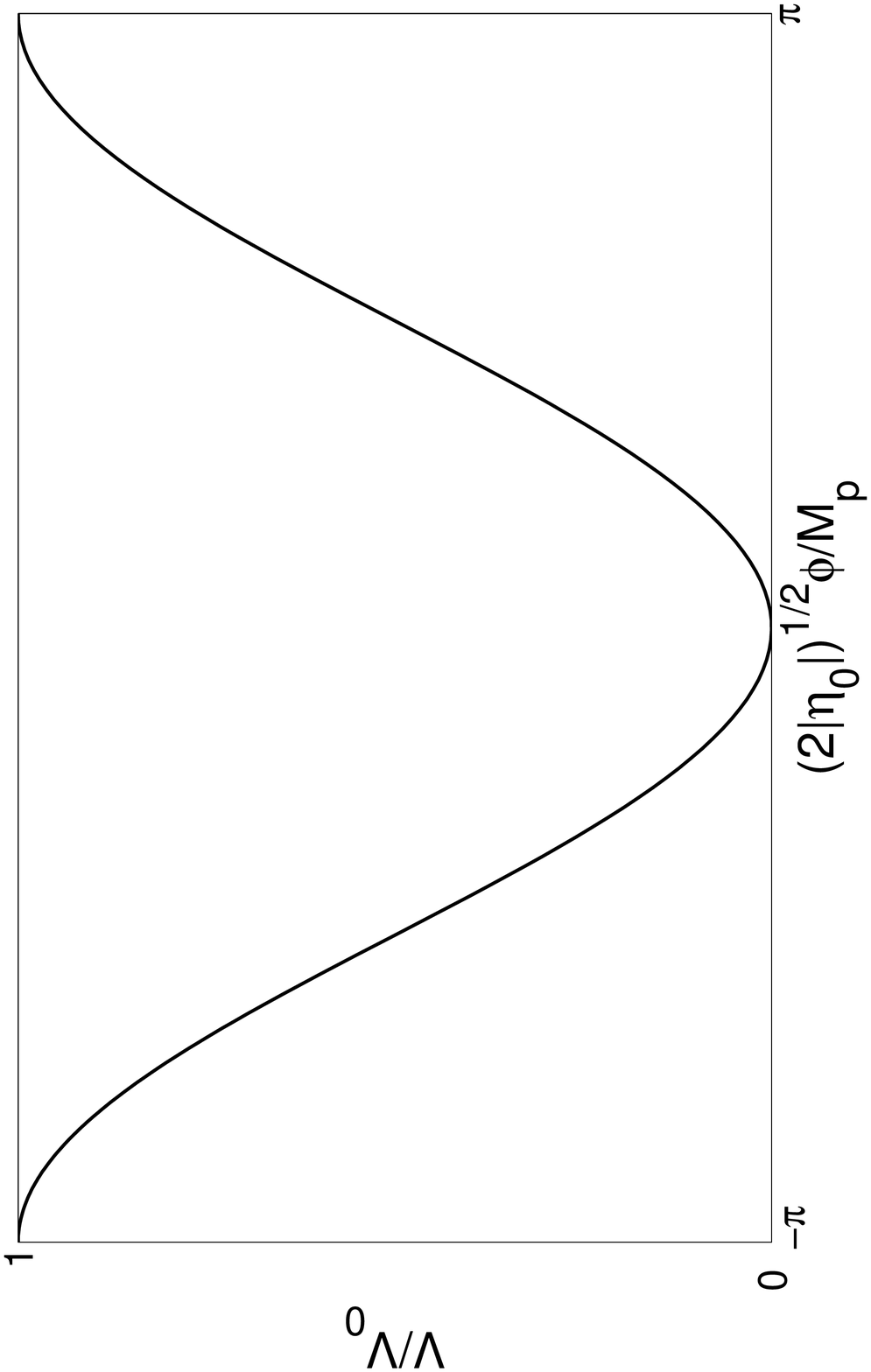}
                                    \caption{Natural/chaotic inflation}
                                    \label{sinusoidal}
   \label{smallv4}
                        \end{minipage}
                        \end{center}
                        \end{figure*}

        We first suppose that inflation takes place near the origin which
is a maximum of the potential,  as
 shown in Figure \ref{smallv1}.
Inflation taking place near a maximum
 is an attractive possibility because eternal
inflation can take place very close to the maximum, providing a natural
initial condition for the subsequent slow roll
\cite{eternal1,eternal2,bl}.

This type of model is usually taken to be non-hybrid,
 so that the vev
 $\vev{\phi}$ is the minimum corresponding to $V=0$, and inflation
ends  with the failure of slow-roll at $\phi\sub{end}\sim
\vev{\phi}< \mpl$. The reason is that one usually thinks of the maximum
as the fixed point of symmetries, which  means that one would be
 dealing
with inverted hybrid inflation \cite{inverted},  as opposed to ordinary
hybrid inflation where the field is moving {\em towards} a fixed point.
Inverted hybrid inflation  is more difficult to arrange than ordinary
hybrid inflation, especially in the context of supersymmetry. On the other
hand, it can be that the origin is not a fixed point, in which case
this type of model can be an ordinary hybrid inflation model.
To keep things simple we take this kind of inflation to be non-hybrid
though the key results still hold if it is hybrid.

We shall take $\phi\sub{end}<\mpl$ to keep within the effective field
theory framework of the last section. Since
$\Delta\phi_N< \phi\sub{end}$ we then deal with a small-field model,
and usually $\Delta\phi_N\sim \phi\sub{end}$.

If $\phiend$ is far below $\mpl$ we shall call this setup New Inflation
 after the first viable inflation model \cite{new}.
If $\phiend$ is not very far below $\mpl$ we will call it modular inflation,
since it is  most plausible realised by invoking the potential
\eqreff{vmod}. Modular inflation was discussed for instance in
 \cite{andrei83,modular1,modular2,graham}.\footnote
{A different way of using moduli to inflate is described in
\cite{glm} but the prediction for the spectral tilt
  depends on details of the model which were not specified.}
If there is more than one modulus, modular
 inflation should take place near a point in the space of the moduli where
the potential
is an extremum with respect to all of them, which may be difficult to arrange.
Also,  the trajectory may lie in the space of two  or more moduli,
corresponding in general to a multi-component inflation model with
non-canonical  normalization. Here we  focus on the case where the trajectory
is mainly in the direction of  one modulus (taken to be canonically normalized)
as for instance in \cite{modular2}. (For an example of the  opposite case
see \cite{ks}.)

{}The potential  \eqreff{vmod} typically gives
 roughly $\epsilon\sim |\eta|\sim 1$, but
the idea is that one gets lucky so that the flatness conditions
\eqreff{flat} are marginally satisfied.\footnote
{In particular, $m^2$ is supposed to be significantly smaller than 
$V_0/\mpl^2$ (see \eq{msquaredofvzero}).
If instead one supposes that $m^2$
 is actually quite a bit {\em bigger} than $V_0/\mpl^2$,  the modulus becomes
a candidate for the waterfall field in  hybrid inflation \cite{supernatural}.
The term
modular inflation is not taken to cover that case.}
 The extent to which
string theory allows modular inflation without fine-tuning
is the subject of intense investigation
at present, as for example in \cite{modular2}.

Considering both New and Modular inflation, let
us suppose that
one term dominates \eqreff{vseries} at least
while cosmological scales are leaving the horizon;
        \be
            V=V_0 \[ 1- \( \frac\phi\mu \)^p + \cdots \]
            \label{vnew}
            \,,
        \ee
We consider first the case that
the dominant term is the leading one;
\bea
            V&=& V_0 - \frac12m^2\phi^2+\cdots  \\
&=&  V_0 \[ 1-  \frac12 \meta \frac{\phi^2}{\mpl^2}
 + \cdots \] \label{vquad}
\,,
\eea
where $\eta_0<0$ is the value of $\eta$ at the maximum.
The spectral tilt is $n-1=2\eta_0$ so that observation
 requires  $\meta\lsim 0.03$.

\begin{figure*}

        \centering\includegraphics[angle=270,totalheight=5in, width=2.0\columnwidth]{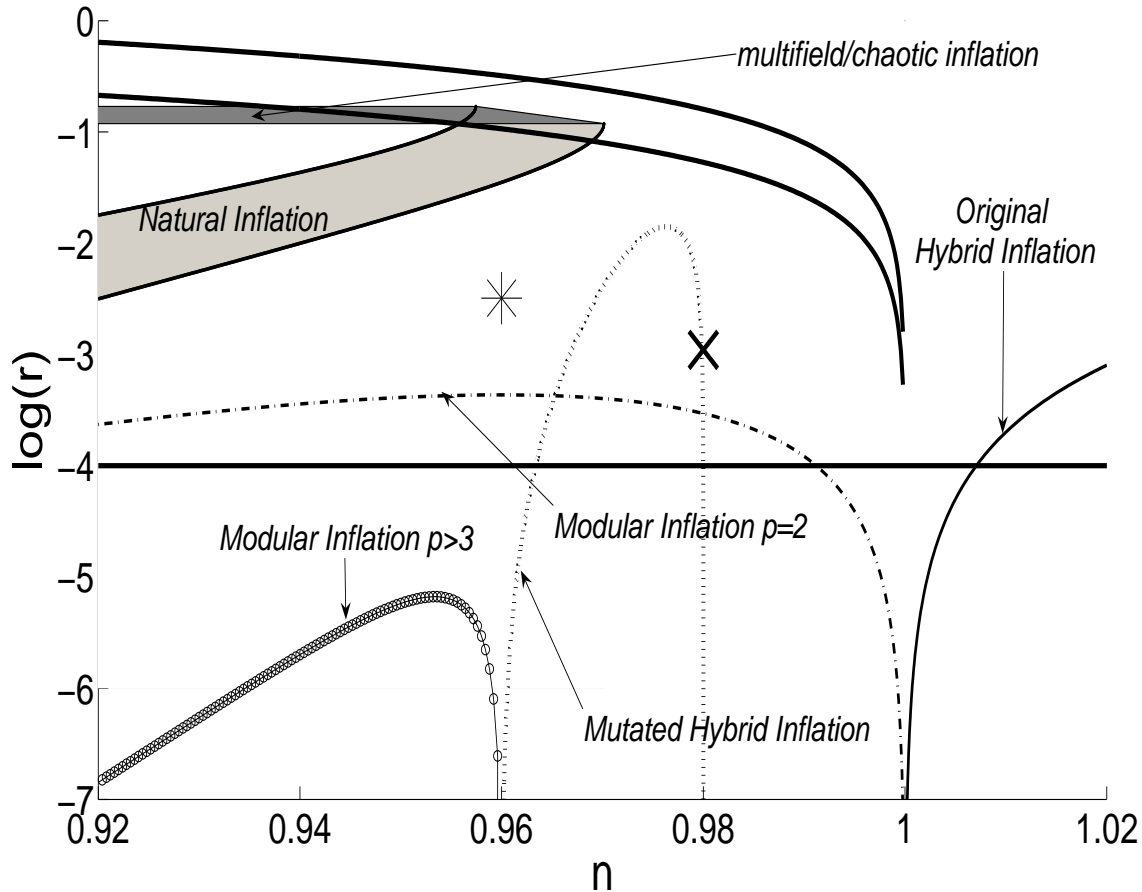}
        \caption{The two upper lines are the ones  shown in  Figures
        \ref{rlog} and \ref{rlinear}, which separate  the three regimes.
        The shaded regions give the
        prediction for natural inflation and multi-component chaotic inflation
        with $N=54\pm7$. Single-component chaotic inflation corresponds
        to the intersection of these two regions. The other predictions
are calculated with $N=50$, and they represent  upper bounds 
obtained by setting $\mu$, $\phi_*$ or $\phi\sub{end}$ 
equal to $\mpl$, as described
in the text. These are the four labeled curves, the cross which corresponds
to the $D$/$F$-term potential \eqreff{loop} 
and the star which corresponds to the
exponential potential \eqreff{vqexp}.
To be eventually observable, $r$ should be above the horizontal
line.} 
%        The line marked  `modular inflation $p=2$' gives the upper bound
%        for the  potential \eqreff{vquad}.
%        The lines marked
%        `modular inflation $p>3$' and
%        `mutated hybrid inflation'  give    upper bounds  on $r$
%        for   inflation potential \eqreff{vnew} with  respectively
%        $p>3$ and $p<0$. The upper bound for the same potential with $2<p<3$
%        will be between these curves and the `modular inflation $p=2$' curves.
%        The line marked `original hybrid inflation' is the upper bound for the original
%        hybrid inflation potential \eqreff{voriginalhybrid}. The star shows the
%        prediction for the exponential potential \eqreff{vqexp}and the cross 
%shows the prediction for $F$-term inflation \eqreff{ftermr}}
                \label{newrvsn}
    \end{figure*}

To achieve a vev
$\vev\phi\lsim \mpl$, the potential \eqreff{vquad}
must steepen sharply after cosmological
scales leave the horizon.
 This would be expected for modular inflation
(though there is no strong reason to expect the quadratic term to
dominate in that case) and it  can be arranged for New Inflation
as in \cite{dr,pseudonat} (see also \cite{steve} for a hybrid
model). As a crude  approximation, suppose that the quadratic term
dominates until inflation ends at $\phi\sub{end}\lsim \mpl$. Then
one finds
\be
\phi_*=\phi\sub{end}\exp(-N(1-n)/2)
\label{phistarofphiend}
\,,
\ee
 and \cite{bl}\footnote
{Even if we allow $\phi\sub{end} \gg \mpl$, one cannot expect the
 potential \eqreff{vquad} to control the end of inflation. This is
because slow-roll inflation with that potential would continue up to
practically $\phi=\vev\phi$. The additional terms generating $\vev\phi$
will surely come in earlier, giving a shape more like the one in Figure
\ref{sinusoidal} that we shall be discussing as a large-field model.}
 \be r = 2 \( \frac{\phi\sub{end}}{\mpl} \) ^2
(1-n)^2  e^{-N(1-n)}
\label{rquad}
\,. \ee
The maximum of this function is below
the bound \eqreff{rbound2} (by a factor $e^{-2}$), consistent with
$\log V$ being concave-downward.The prediction $r(n)$ is plotted
in Figure \ref{newrvsn} for
 $\phi\sub{end}=\mpl$  and $N=50$. Since $\phi\sub{end}$ is actually
expected to be significantly smaller we conclude that $r$ is 
unlikely to be observable.

This model, with the quadratic term dominating,
is reasonable only if $n\lsim 1-1/N\lsim 0.98$. Otherwise,
from  \eq{phistarofphiend},
the quadratic term would have to dominate higher-order terms,
up until practically the point where those
 terms become so important that they steepen the potential causing
an immediate end to inflation. Such an abrupt transition is clearly
unreasonable in the context of non-hybrid inflation.

\begin{figure}
 \centering\includegraphics[angle=270, width=1.0\columnwidth,totalheight=2.5in]{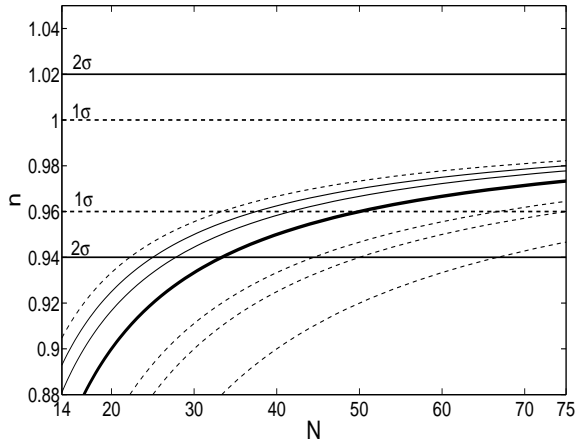}
               \caption{The prediction  \eqreff{nnew} for different $p$.
                The bold full line is the limit $|p|\to\infty$. Above it from top down are the
                lines $p=0$, $-2$  and $-4$, and below it from bottom up are the lines
                $p=3$, $4$ and $5$. The observational
bounds from \cite{sdsswmap} are indicated.}
               \label{nvsN}
            \end{figure}

All of this assumes that the quadratic term  dominates  when cosmological
scales leave the horizon. The situation is quite different
if we assume instead that a higher-order term has become  dominant
by that time. This case has often been considered  as
for instance in \cite{moreusual}, and for modular inflation in
\cite{graham}.
% it is supposed that
%the bound on $\meta$ makes  the quadratic term negligible
%by the time that cosmological scales leave the horizon
Adopting it,
 we arrive at
\eq{vnew} with $p\geq 3$.
Inflation ends at $\phi\sub{end}\lsim \mu$, and
to have a small-field model we will take $\mu\lsim \mpl$. Then
\bea
N&=& \frac1{p(p-2)} \( \frac\mu\mpl \)^2
\(\frac\mu{\phi_*}\)^{p-2} \[ 1 -
\(\frac{\phi_*}{\phi\sub{end}} \)^{p-2} \] \label{Nnew0} \nonumber \\
&\simeq&
 \frac1{p(p-2)} \( \frac\mu\mpl \)^2
\(\frac\mu{\phi_*} \)^{p-2} \label{Nnew}
\,.
\eea
The final equality holds because we are requiring $\phi_*/\phi\sub{end}$
appreciably less than 1, for the reason discussed earlier in connection
with the quadratic potential.
This gives, independently of $\mu$,
        \be
            n   = 1 - \frac {p-1}{p-2} \frac 2{N}
            \label{nnew}
            \,.
        \ee
The absolute constraint $N<75$ implies  $n<0.987$.
This prediction is plotted against $N$
in Figure \ref{nvsN} for $p=3$, $4$ and the limiting case
        $p\to\infty$.
We see that $p=3$ is disfavoured by observation, and that for all $p$
observation provides  a significant lower bound on $N$.
 Adopting the reasonable
range \eqreff{Nest3} the prediction  becomes
\be n = 1 -( 0.037 \pm 0.005) \frac{p-1}{p-2}
\,. \ee
This prediction is in the range $n<0.967$.

\begin{figure}
                \centering\includegraphics[angle=270,width=1.0\columnwidth,totalheight=2.5in]{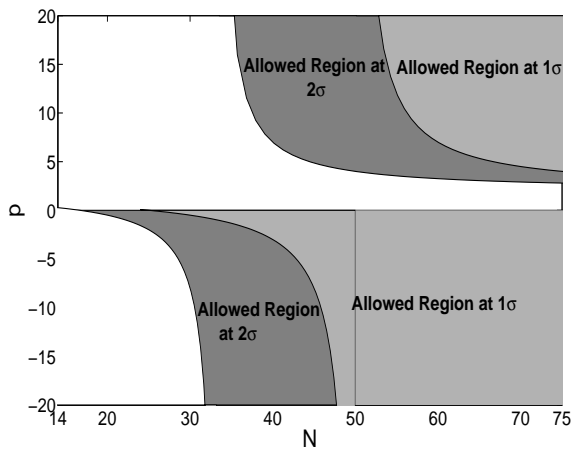}
                 \caption{The regions excluded by the observational bounds
from \cite{sdsswmap}, for the parameter $p$ in the prediction \eqreff{nnew}.}
                 \label{pvsN}
            \end{figure}

\begin{figure}
                \centering\includegraphics[angle=270,
                width=1.0\columnwidth, totalheight=2in]{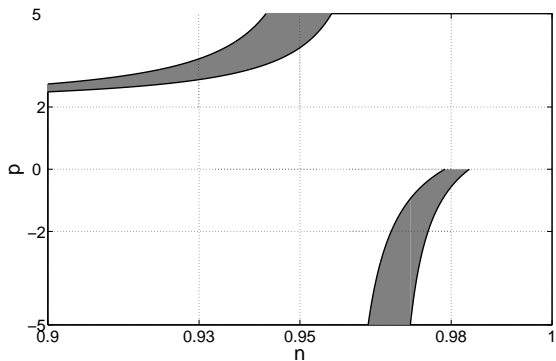}
                \caption{The prediction \eqreff{nnew}
for $N=54\pm 7$.}
                \label{pvsn}
            \end{figure}

We have been assuming that a single power in \eq{vnew} dominates, which may be
unreasonable  for modular inflation. But even
if a single power does not dominate, \eq{vnew} could well be a useful
        approximation for the relevant values of $\phi$, with some
non-integral $p$.
%        Then,  if $p$ is significantly bigger than  2, \eq{nnew}
%        will hold. Such a description plausibly applies  to
%modular inflation with $\mu\sim\mpl$.
 Certainly,
the specific models of \cite{modular2} give the small negative tilt
and the  high inflation scale which is the characteristic of that
approximation.
Treating $p>3$ as a continuous variable, the
 allowed region in the
        $p$-$N$ plane is shown in Figure \ref{pvsN}.
Taking $N=54\pm 7$, we plot $n(p)$ in  Figure \ref{pvsn}.

%    \begin{figure*}
 %   \begin{minipage}[t]{0.5\linewidth}
  %  \centering\includegraphics[angle=270, width=3in]{compare_r_new.eps}
   % \caption{The sensitivity of the modular inflation tensor fraction
%to $\mu$. {}From the top the curves correspond to
%$\frac{\mu}{M_{p}}=10,8\pi,1,0.1$.}
 %   \label{compare_r_new}
  %  \end{minipage}%
%   \begin{minipage}[t]{0.5\linewidth}
%    \centering\includegraphics[angle=270, width=3in,totalheight=2in]{compare_r_new_bigger.eps}
%    \caption{figure(\ref{compare_r_new}) for a bigger range of $n$ to illustrate the sensitivity of the local minima to $\frac{\mu}{M_{p}}$}
%    \label{bigX}
%    \end{minipage}
% \end{figure*}

The  cmb normalization \eqreff{cmbnorm} corresponds to a
tensor fraction
     \be r = 8\left(\frac{\mu}{M_p} \right)^{\frac{2p}{p-2}}
p^2(p(p-2))^{-2\left(\frac{p-1}{p-2}\right)}
N^{-2\left(\frac{p-1}{p-2}\right)}
\label{rnew}
      \,. \ee
Taking $N=50$,  this
  prediction is plotted in Figure \ref{newrvsn}
for $\mu=\mpl$.
% and in Figure \ref{compare_r_new}
% for $\mu=10\mpl$  $\mpl$ and $0.1\mpl$.
We see that $r$ is too small ever to observe.

In all of this we focused on the case $p>3$. Allowing instead $p-2$
to be quite small, the first line of \eq{Nnew} should generally be used.
Then, taking $\phi\sub{end}\simeq \mpl$,
the prediction for $r(n)$  will interpolate between the $p=2$ and $p>3$
 curves in Figure \ref{newrvsn}. 
 As with the quadratic potential, the
model makes sense (as a non-hybrid one) only if $n\lsim 0.98$ since
otherwise $\phi\sub{end}/\phi_*$ would be very close to 1.

\subsection{$F$- and $D$-term inflation}
\label{df}
In a supergravity theory the potential  has two parts, called the
$F$ term and the $D$ term. They are  constructed from three functions,
called the superpotential, the Kahler potential and the gauge-kinetic
function.  There is also a possible constant contribution to the $D$-term
called a Fayet-Iliopoulos term.\footnote
{We invoke at most one  gauge-kinetic function and one Fayet-Iliopoulos
term, those quantities appearing only in the $D$-term inflation model.}
 With the simplest (minimal) Kahler potential and
gauge-kinetic function, simple forms for the superpotential can be written
down   which at tree-level are perfectly flat, and can give either
$F$-term  \cite{cllw} or $D$-term \cite{ewanexp} inflation.
The   loop correction then dominates \cite{dss,dterm}, leading to a
hybrid inflation model.
To ensure that
the one-loop correction
is  a good approximation the renormalization scale $Q$ should be
chosen to be of order $\phi$.
%Since $Q$ must also be at most of order the ultra-violet cutoff,
%this makes mandatory our requirement $\phi\lsim\mpl$.

In a non-supersymmetric theory the loop correction would generate
a term
$\lambda\phi^4\ln(\phi/Q)$; it was invoked in the original New Inflation
model \cite{new} and it practically corresponds to \eqreff{vnew} with $p=4$.
In a globally supersymmetric theory with soft supersymmetry breaking,
the loop correction corresponds  to a running mass
term $m^2\phi^2\ln(\phi/Q)$. Taken as an approximation to  supergravity,
this gives a  running-mass model of inflation \cite{ewanrunning},
 whose observational signature is discussed elsewhere
\cite{cl,ourrunning}.

 The case at hand corresponds (as an approximation to supergravity)
to a globally supersymmetric theory with
spontaneous supersymmetry breaking, with the
loop correction coming
 from the waterfall field in a hybrid inflation model.
Assuming   $\phi\gg\phi\sub{end}$, the  potential is
        \be
            V = V_0 \[ 1 + \frac{g^2}{8\pi^2}\ln \frac\phi Q \]
            \label{loop}
            \,,
        \ee
where $g\lsim 1$ is the coupling of $\phi$ to the waterfall field.
        {}With $Q\simeq\phi$, $V_0$ dominates and
        $V'/V = g^2/8\pi^2\phi$. The shape of this potential
is  illustrated in Figure \ref{smallv2}. The integral
\eq{nofk} is again be dominated by the endpoint
        $\phi_*$, so that the tilt
 is given by \eq{nnew} with $p=0$
as $n-1=-1/N$.
 The observational bound is shown in
        Figure \ref{nvsN}.
For this case, the normalization \eqreff{cmbnorm} gives a  tensor fraction
\be
r= \frac1{2\pi^2} \frac{g^2}{N} = 0.0011 \( \frac{50}{N} \)^2 g^2
\label{ftermr}
\,.
\ee
For $g=1$ this prediction is shown in Figure \ref{newrvsn}.
We  see that $r$ will never be observable if $g$ is much below 1.

Although these predictions are clean, they may not be realistic.
Even within the setup we have described, there is a regime of parameter
space where $\phi$ is close to $\phi\sub{end}$ so that
 the coefficient  of the log in
\eq{loop} is reduced \cite{reduce}.
Much more seriously, the
 potential \eqreff{loop} gives
\be
\phi_*  = \sqrt{\frac N{4\pi^2} } g \mpl
\label{phisg}
\,.
\ee
Unless the coupling is very small, this is only marginally a small-field
model. A very small coupling is reasonable for the $F$-term case, but not
for the $D$-term case.

If the coupling is not very small,
 the minimal
forms for the Kahler and gauge-kinetic functions cannot be justified at all
\cite{cllw,mydterm}, while the simple form for the superpotential requires
an almost exact global symmetry which is usually considered implausible.
For these reasons, the   tree-level potential is unlikely to be
precisely flat in that case.
Including it could significantly alter the shape of the
potential \cite{reduce}, possibly generating  a maximum \cite{bl,sy}.
In that case, after
 re-defining the origin as
the maximum, \eq{vnew} may be an adequate approximate for
some effective $p\gsim 3$. In any case, one expects that these
$F$- and $D$-term inflation models will give $n<1$ corresponding to a
concave-downward potential.

Before leaving $F$- and $D$-term inflation, we remark on their theoretical
status. In both cases, the scale of inflation is related to the vev $\Lambda$
of the waterfall field by $V_0=g^2\Lambda^4$, and the cmb normalization
determines the latter independently of $g$;
\be
\Lambda = \( \frac{50} N \)\half \times 6 \times 10^{15}\GeV
\,.
\ee
(This is equivalent to the fact that \eq{ftermr} contains a factor $g^2$.)
In the case of $F$-term inflation, the waterfall field can be identified with
a GUT Higgs field. Then $\Lambda$ should be essentially the
unification scale deduced from observation, which is $2$ to $3\times
10^{16}\GeV$. Considering the theoretical uncertainties the agreement is
satisfactory, making  $F$-term inflation one of the few inflation
models which relate  the observed
value of $\calp_\zeta$ to parameters of particle
physics. In the case of  $D$-term inflation, $\Lambda$ is the normalization
of the Fayet-Iliopoulos term. It is related to
the string energy scale, but the relation depends on the version of string
theory that is relevant \cite{rr,dreview}.

\subsection{More concave-downward potentials}
\label{concdown}
Next we consider the potential \eqreff{vnew}, but with $p<0$ to give the
shape of Figure \ref{smallv2}. Within the context of the setup described in
Section \ref{effective}, this potential can be generated by
  mutated hybrid inflation
\cite{mutated,smooth,inverted}. In mutated hybrid inflation
 the potential is supposed to have negligible variation in the $\phi$
direction with all other fields fixed at the origin, but there is supposed
to be a coupling with some heavy field.
At each $\phi$ this field is fixed
at the instantaneous minimum of its potential (integrated out), so that the
potential $V(\phi)$ during inflation in fact has significant variation.
The potential typically is of the form \eqreff{vnew} with $p<0$,
 with   integral values of $|p|$ favoured but not mandatory.\footnote
 {Mutated hybrid inflation
  can also give \eqreff{vnew} with $p>1$ though less typically.
The assumption that the potential  has negligible variation in
the $\phi$ direction is
not mandatory. Such variation could generate a maximum for the potential,
giving a situation similar to the one that
we described  after \eq{phisg}.}
With integral $p<0$, a  potential of the form \eqreff{vnew}
 has also been suggested in $N=2$ supergravity
\cite{juann2} and in  D-brane cosmology  \cite{dbraneinf}.

The predictions are given by the same formulas \eqsref{nnew}{rnew}
as for $p>2$,  taking  again  $\phi$ and $\phi\sub{end}$ to be sufficiently
well-separated. The  allowed region in the $p$-$N$ plane is shown
        in Figure \ref{pvsN}. The prediction  $n(N)$ is
 shown in Figure \ref{nvsN} for
 $p=-1$, $-2$ and $-3$.  The observations do not constrain $p$, but they do
constrain $N$ significantly.  For $N=54\pm7$ the prediction $n(p)$ is
shown in Figure \ref{pvsn}. The  prediction $r(n)$ with $\mu=\mpl$ is
shown in Figure \ref{newrvsn}.

Another concave-downward potential is
\be V\simeq V_0(1-e^{-q\phi/\mpl}) \label{vqexp} \,.
\ee
This potential has the shape in Figure \ref{dbranes}, and it
may be
generated by  a kinetic term passing through zero \cite{ewanexp},
giving $q=\sqrt2$. The same form can arise in non-Einstein gravity
inflation \cite{treview} with $q=\sqrt{2/3}$. This case
is actually on the border of the small-field regime, corresponding
to $\Delta\phi_N\sim \mpl$, but the  kinetic term
and the non-Einstein
mechanisms both fall outside the framework we set out at in Section
\ref{effective}.  The prediction for
the spectral tilt  is given by \eq{nnew} with $p=\infty$ as $n-1=-2/N$.  The
prediction for $r$ is \be r=\frac 8{q^2 N^2} \sim \frac 8{N^2} \,,
\ee which should be observable as shown in Figure \ref{newrvsn}.
 A similar form has been derived for the potential of a
modulus \cite{cq}, which gives  the same prediction for $n$ but
negligible $r<10^{-9}$.

\subsection{Concave-upward potentials}
\label{concup}

Next we consider the original hybrid inflation potential \cite{originalhybrid},
\bea
V&=& V_0 + \frac12 m^2\phi^2\\
&=& V_0 \( 1 + \frac12 \eta_0\frac{\phi^2}{\mpl^2} \)
\label{voriginalhybrid}
\,,
\eea
It is supposed that $\eta_0\ll 1$,  so that
inflation ends only when the waterfall field responsible for $V_0$
is destabilized. The constant term $V_0$ dominates for $\phi/\mpl
\ll \eta_0\mone$.
In that regime, $\phi_*=\phiend \exp[\frac12 (n-1)N]$.
This gives
\be
n-1=2\eta_0>0
\,.
\ee
which might be indistinguishable from $1$.
The tensor fraction is
\be
r=2 \(\frac{\phi_*}\mpl \)^2 (n-1)^2
\,.
\ee
Working within the framework of Section \ref{effective} we need
$\phi_*<\mpl$. As shown
in Figure \ref{newrvsn}, the tensor fraction might be observable within this
regime.

The original hybrid inflation model
 makes   good contact with
field theory beyond the Standard Model. In contrast with the other potentials
we consider, the inflation scale $V_0$ can be quite low, which would reduce
 $N$. (Happily, the predictions for the spectral tilt is 
 independent of $N$.) 
In the context of supersymmetry $V_0\quarter\sim 10^{10}\GeV$
is a natural choice \cite{supernatural},
with $\eta_0$ not too small so that $m\sim \TeV$.
The
 parameter space model is limited however, by the requirement
that the loop correction from the waterfall field be negligible
\cite{myhybrid}. Supersymmetric hybrid inflation of any kind
is impossible with $V_0\quarter\sim \TeV$ which is
desirable for instance  \cite{mytev,kl}
in the presence of  extra dimensions. (See however \cite{clrt}
for a non-supersymmetric hybrid inflation model with $V\quarter\sim\TeV$,
which can generate baryon number as well.)

Next we consider the potential
        \be
            V=V_0 \[ 1 + \( \frac\phi\mu \)^p + \cdots \]
            \label{vstrange}
            \,,
        \ee
with the constant term dominating and $p\geq 2$ or $p<0$.
For these values of $p$ the   potential has  the
shape shown respectively in Figures \ref{smallv3} and \ref{smallv4}.
An integer $p\geq 3$ would correspond
 \cite{andreihybrid,hybrid2} to a higher-order term dominating,
 and non-integral values $p>2$ can be motivated as an approximation
(cf.\ the discussion in connection with \eq{vnew}) or possibly from mutated
hybrid inflation. Negative integral values of
$p$ could correspond to dynamical supersymmetry breaking \cite{dsb},
or to $D$-brane cosmology \cite{dbraneinf}.

In all of these  cases, inflation can occur only in a
limited region $\phi<\phi\sub{max}$, where
 $\phi\sub{max}$ corresponds to  $\eta\sim 1$, and it is not clear how the
field is supposed to arrive in that region.
The integral  \eq{nofk} is dominated by the endpoint $\phi\sub{end}$
if it is sufficiently well separated from $\phi_*$. Then the spectral tilt
and tensor fraction are given by \eqs{nnew}{rnew}, with the replacement
$N\to N\sub{max}- N$
where $N\sub{max}$ is the maximum number of $e$-folds. Barring fine-tuning
one  expects  $N\sub{max}\gg N$, which probably will make the spectral
tilt and the tensor fraction too small to observe.

There remains the case of potentials \eqsref{vnew}{vstrange} in
the range $0<p<2$, which looks unlikely in the context of field
theory. For the record, 
%the  following statements apply.
%In the range  $0<p<1$, 
%$V$ and $\log V$ are concave-downward for for the potential \eqreff{vstrange} 
%and
%concave-upward for $p$ outside that range excluding the endpoints.
%In the range  $0<p<1$, 
%$V$ and $\log V$ are concave-upward  for for the potential \eqreff{vstrange} 
%and
%concave-downward for $p$ outside that range excluding the endpoints.
% For the potential \eqreff{vnew} each
%of these alternatives is exchanged.
 if $\phi\sub{end}$ and $\phi$ are sufficiently well
separated, \eqsref{nnew}{rnew} apply for  the potential
\eqreff{vstrange}, while for the potential \eqreff{vnew} these equations
apply with $N\to N\sub{max}-N$.

        \section{Single-component Natural/Chaotic  inflation}\label{natural}

\label{s:naturalchaotic}

\subsection{The single-component case}

Now we consider large-field models, corresponding
to  $\Delta\phi_N> \mpl$. These lie outside the effective field theory
framework described in Section \ref{effective}.

Most discussions of the large-field case
adopt a `chaotic inflation'
potential
\be
V\propto \phi^{p}
\label{vchaotic}
\,,
\ee
 with $p$ an even integer. (The term `chaotic'
is used because the potential was first introduced \cite{chaotic}
as the simplest one which would support what Linde called a chaotic
initial condition.) With such a potential,
 $\phi_* \simeq \sqrt{2Np} \mpl$, and
$\phi\sub{end}\sim \mpl$. With this form of the potential $V'/V
\propto 1/\phi$ making $\log V$ concave-downward. The integral
\eqreff{nofk} is  practically independent of $\phi\sub{end}$
leading to the predictions \bea
n&=& -   \frac{2+p}{2N} \\
r &=&  \frac{4p}{N}
\,.
\eea
Observation rules out $p\geq 4$, but marginally allows
$p=2$.
In Figure \ref{rvsn} we  show the prediction for $p=2$, with
$20<N<75$.
Using the reasonable  range \eqreff{Nest3} it becomes
\bea
n&=& -0.037 \pm 0.005 \\
r&=& 0.15 \pm 0.02
\eea
 We show this  prediction in Figure \ref{newrvsn}.

Before continuing we address the following point.
In a
non-supersymmetric theory, where $V$ is part of the lagrangian, one is free
to specify the potential
\eqreff{vchaotic} even though it goes beyond the effective
field theory framework of Section \ref{effective}.
But in  the context of  supergravity, where $V$ (taking it to come from the
$F$ term) has to be constructed from the
 Kahler potential and the superpotential one might wonder whether a given
form is possible at all.
That this is so has been demonstrated for $V\propto \phi^2$
by writing down explicit forms for the Kahler potential and superpotential
 \cite{adhoc,kyy,yy}, and similar constructions would surely work
for any power. Thus, the status of the
potential $V\propto\phi^p$ in supergravity is the same as
  in a
non-supersymmetric field theory.\footnote
{The supergravity  proposals of \cite{kyy,yy} declare that there is a
  shift symmetry $\phi\to\phi+$\,const, which is
 broken by a specific   superpotential chosen  to give
$V=\frac12m^2\phi^2$ (up to a small correction).
This is conceptually the same as declaring, in a
non-supersymmetric theory, that there is a shift symmetry broken only
by $V=\frac12m^2\phi^2$ itself;
in both cases a similar  declaration could be made
to justify any desired potential (which in the case of supergravity can be
constructed from a Kahler potential and a superpotential). The supergravity
proposals  of \cite{kyy,yy} have an
 additional complication though, that the chosen form of the superpotential
corresponds to an exact global $U(1)$ symmetry acting on a field different
from $\phi$. That symmetry presumably is broken and it is not clear how the
breaking might affect the model.}

The
 potential $V\propto \phi^2$ may be   better motivated if it is
considered as an approximation
to a sinusoidal potential near a minimum. Such a potential was
 dubbed Natural
 Inflation by the people who first considered it \cite{natural}.\footnote
{The potential is periodic because $\phi$ is supposed to be a PNGB.
The Natural Inflation potential is flat enough for inflation, only because
 the period is taken to be much bigger than $\mpl$. The small-field
PNGB models mentioned earlier \cite{ewanmulti,pseudonat} use different
 potentials.}
To achieve inflation the period of the potential has to be much bigger than
$\mpl$, but proposals have been made
\cite{extranat,pseudonat,Kim:2004rp}
which motivate
 such a large period.\footnote
 {In \cite{bst} it is stated that these proposals generate literally
a potential
        $V\propto\phi^2$, corresponding (with fixed $N$)
to a point in the $r$-$n$ plane. In fact the sinusoidal potential corresponds
        to a line as we are about to see.} The first two of these go beyond
the framework of Section \ref{effective} by invoking non-trivial
extra dimensions, while
the last stays within it by making the field $\phi$ correspond to a path
in field space which winds many times around the fixed point.

The sinusoidal potential can be written
        \be
            V=  V_0 \sin^2(\sqrt{\meta/2} \phi/\mpl)
            \label{nat}
            \,,
        \ee
and is plotted in Figure \ref{sinusoidal}.
Provided that $\meta\ll 1$, the potential supports inflation until
 $\eta\sim\epsilon\sim 1$. {}From   \eq{nofk},
        \be
\sin\( \sqrt{
\frac { \eta_0}2
}
\frac{ \phi_*}\mpl \)\simeq
\sqrt{
\frac 1{1 + \eta_0}
}
\,e^{-N \eta_0}
\,,
        \ee
leading to
        \bea
            \epsilon &=&  \frac1{2N}
\frac{2N\meta}{e^{2N\meta}-1} \label{naturaleps} \\
            \eta &=&  \epsilon -  \meta
            \,.
        \eea

{}From these expressions one can read off $n-1$ and $r$. At fixed
$N$ the prediction depends on the parameter $|\eta_0|$. The
results for $N=20$ and $N=75$ are shown in
 Figure      \ref{rvsn}, and the result for the reasonable range
\eqreff{Nest3} is shown in Figure \ref{newrvsn}. We see in the latter
figure that as $n$ goes more negative we move from  the regime
$V''(\phi_*)>0$ into the regime  $V''(\phi_*)<0$. Note though that the
requirement $|\eta_0|\ll 1$ always makes this a large-field model
($\Delta\phi_N>0$); we see again that the labeling of the left-hand
region in Figures \ref{rlog} and \ref{rlinear}
as `small-field' is inappropriate.

\section{Multi-component Chaotic  Inflation}
\label{s:multi}

So far in this paper we dealt only with
 single-component slow-roll
inflation, where the inflationary trajectory is
essentially unique; it  is either the only possible one in the space of the
scalar fields, or else is one of a family of straight-line trajectories
which are equivalent during inflation. An important example of the latter
case is a  potential  $V(\phi)$, with
\be
\phi^2= \sum_{i=1}^K  \phi_i^2
\label{phiofphii}
\,.
\ee
We take the $\phi_i$ to be canonically normalized, making the radial field
 $\phi$ also canonically normalized along each direction.
Each direction corresponds to a possible inflationary trajectory, but
an  $SO(K)$  symmetry leaves $\phi^2$ invariant
and transforms the  trajectories into each other.

 For a multi-component slow-roll model, there is a family of possible curved
trajectories in the space of two or more fields, which we refer to as
components of the inflaton. The
potential is supposed to satisfy flatness conditions analogous
to \eq{flat}
\be
\epsilon_i \ll 1 \qquad |\eta_{ij}|\ll 1
\,,
\ee
where
\bea
\epsilon_i &\equiv & \frac12 \mpl^2 \( \frac{\pa V/\pa \phi_i}{V} \)^2 \\
\eta_{ij} &\equiv& \mpl^2 \frac{\partial^2V/\pa\phi_i\pa\phi_j}{V}
\,.
\eea
The field equation for each field $\phi_i$ is supposed
to be well-approximated by $3H\dot\phi_i=-\pa V/\pa \phi_i$,
so that the possible inflationary
trajectories are the lines of steepest descent of the potential.
The set of all
fields satisfying these conditions may be called the set of light
fields.

While in principle every field satisfying these conditions could
be regarded as components of the inflaton, in practice one will
include only those fields corresponding to directions in which the
slope $\pa V/\pa \phi_i$ is big enough to lead to significant
curvature. Light fields corresponding to a smaller slope will have
no significant effect during inflation, though they may come into
play afterward and be the main source of the curvature
perturbation.

To evaluate the curvature perturbation generated by the vacuum fluctuations
of the multi-component inflaton, we can use the
$\delta N$ formalism  \cite{starob85,ss,lms,lr,st} which handles all light
fields on an equal footing whether or not they are significant during
inflation.\footnote
{For the present purpose, which is to evaluate the curvature perturbation at
the end of multi-component inflation, one could instead use perturbation
theory which reduces the problem to the solution of linear equations.
The equations are well-known at first order
(see \cite{btw} for a review) and progress has recently been made towards their
extension to second order \cite{karim}. Another formalism for multi-component
inflation is presented in \cite{rst}. For multi-component chaotic inflation
within the slow-roll approximation
 the $\delta N$ formalism is preferable, because it gives simple
expressions which are valid at both first and second order in perturbation
theory.} Keeping quadratic terms \cite{lr}, the
time-dependent curvature perturbation smoothed on a given comoving scale
$a/k$ is
\bea
\zeta(\bfx,t)  &=& \delta N(k,\phi_i(\bfx),\rho(t)) \nonumber\\
&=& \sum_i N_i \delta\phi_i(\bfx) + \frac12 \sum_{ij}
N_{ij} \delta\phi_i \delta\phi_j
\label{zetaofdphi}
\,.
\eea
Here, $N(k,\phi_i,\rho)$ is the number of $e$-folds, evaluated in an
unperturbed universe, from an epoch soon  after the smoothing scale leaves the
horizon when  the fields $\phi_i$ have specified values,
 to an  epoch when the energy density $\rho$ has a specified
value.
In the  second line, $N_i\equiv \partial N/\partial \phi_i$
and  $N_{ij}\equiv \partial^2 N/\partial \phi_i\phi_j$, both evaluated on
the unperturbed trajectory. In  known
cases  the first two terms of this expansion in the field perturbations
are  enough.

These expressions involve
the unperturbed inflationary
trajectory. It  is
not determined by the potential during observable inflation, if
 more than one  field is light then.
It may be that well before
the observable Universe leaves the horizon there is only one relevant light
field, leading to an essentially unique trajectory at the classical level.
In any case we  suppose that  somehow the unperturbed trajectory
during observable inflation is known.

To evaluate a given Fourier component of $\zeta(\bfx,t)$ we can adopt a
smoothing scale just a bit shorter than the inverse wavenumber.
Thus we may in practice identify $k$ in the above expressions with the
wavenumber of the Fourier component.

The field perturbations $\delta\phi_i$ are generated from the vacuum
fluctuation. They are practically gaussian and uncorrelated, and each
of them has the  spectrum $(H_k/2\pi)^2$, where the subscript $k$
denotes the epoch of horizon exit $k=aH$.
At some stage before nucleosynthesis,  $\zeta$ settles down to a
time-independent value, which is constrained by observation.
We write down the predictions for $\zeta$, which follow from \eq{zetaofdphi}
at any epoch
 even though we are interested in the regime where $\zeta$
has settled down to its final value.

Since the observed $\zeta$ is almost gaussian, one or more linear terms
must dominate \eq{zetaofdphi} at least eventually, giving the spectrum
\be
\calp_\zeta(k)  = \sum N^2_i(k)  (H_k/2\pi)^2
\label{multispec}
\,.
\ee
 Using the multi-component slow-roll formalism the
 spectral tilt is found to be \cite{ss,treview}
\be
n-1 = -2\epsilon -\frac2{\mpl^2 N_iN_i} + 2\frac{\eta_{jk} N_jN_k}{N_mN_m}
\,,
\label{multitilt}
\,.
\ee
where   identical indices are summed over and $\epsilon=\sum_i\epsilon_i$
is given by \eq{epsilondef} with $V'$ the gradient of $V$.

The field basis at horizon exit can be chosen so that one field
$\phi$ points along the inflationary trajectory. Its contribution to
\eq{multispec} is time-independent, and if it dominates  the final value
one recovers the  predictions \eqssref{spec}{ninf0}{rofv}.
The other contributions are initially negligible (almost-exponential inflation
being assumed at horizon exit) but may grow to become significant or dominant.
The tensor perturbation on the other hand depends only on $H_k$ and is
time-independent, which means \cite{ss,treview,book}
that the tensor fraction cannot exceed the prediction \eqreff{rofepsilon};
\be
r \leq 16\epsilon
\,.
\ee

If the contribution of  $\phi$ dominates, the non-gaussianity of $\zeta$
is  too small  to ever be measurable \cite{maldacena,sl}. Otherwise
it may be observable. The likely observables are
 the bispectrum and trispectrum, which alone are generated by the quadratic
expansion \eqreff{phiofphii}. They are specified respectively by
\cite{bl} quantities
$\fnl$ and $\tnl$. Taking the field perturbations to be perfectly gaussian,
which has been justified for the bispectrum \cite{sl,lz}, and ignoring
the scale-dependence of the spectra, the predictions
are \cite{lr,bl2}\footnote
{The general
formula for $\tnl$ follows from the special cases
in \cite{bl2} but has not been written down before. The scale-dependence
of the spectra is included for $\fnl$ in \cite{dg}.}
\bea
-\frac35\fnl &=&
\frac{  N_i N_{ij} N_j }{2  \(  N_m  N_m  \)^2 }
+ 4A \calp_\zeta \frac{ {\rm Tr\,} N^3 }{ \(  N_n N_n  \)^3 }
\label{fnl} \\
\tnl &=& 2 \frac{  N_i N_{ij} N_{jk} N_k }{ \(  N_m N_m \)^3 }
+ 16B \calp_\zeta \frac{ {\rm Tr\,} N^4 }{ \(  N_n N_n  \)^4 }
\label{tnl}
\,,
\eea
where $A$ and $B$ are of order 1 on cosmological scales.
Present observation \cite{fnlbound}
gives roughly $|\fnl|\lsim 100$, and absent a detection the eventual
bound will be \cite{bartolorev} $|\fnl|\lsim 1$. There is at present
no bound on $\tnl$ from modern date, and no estimate of the bound that will
eventually be possible. (A crude bound from COBE data \cite{bl}
 is $|\tnl|<10^{8}$.)

The shape of the
potential of a multi-component inflation model is constrained by these
predictions, if $\zeta$ is assumed to have reached its final value by the
end of inflation.
Stewart and collaborators have exhibited some small-field multi-component
models \cite{ewanmulti,ks}, along with their prediction for the spectral tilt
which depends strongly on the parameters of the potential.
We are going to consider instead
multi-component chaotic inflation, corresponding to
\be
V= \frac12 \sum_{k=1}^K m_i^2 \phi_i^2
\label{multiv}
\,.
\ee
This  potential  gives uninterrupted  slow-roll inflation if the masses are
not too unequal \cite{interrupt}, and we assume that this is the case.

If the field values are of the same
order, inflation ends when
$\phi_i\sim \mpl/\sqrt K \ll \mpl$, which can be much less than $\mpl$
if there are many fields. Recalling  the discussion of
Section \ref{effective}, one might hope that the potential in that case is
generically under good control, with non-renormalizable terms negligible.
This  general idea was termed Assisted Inflation by those who first considered
it  \cite{assisted}.
As the  authors of \cite{Nflation} point out,  whether this is so
should be considered case by case. One can see this by considering
the  case of equal masses, $m_i^2=m^2$, which
gives
\be
V= \frac12 m^2\phi^2
\ee
with $\phi^2$ given by \eq{phiofphii} and  inflation taking  place at
$\phi\gg\mpl$. If this potential were exact, there would be a
$SO(K)$ symmetry which would forbid the appearance of non-renormalizable
terms $\lambda_d\phi_i^d/\mpl^{d-4}$; instead there would be
terms  $\lambda_d\phi^d/\mpl^{d-4}$ which presumably would spoil
inflation at $\phi\gg \mpl$ even though the individual $\phi_i$ can all be
small.

As a specific way of keeping the multi-chaotic potential under
control, the authors of \cite{Nflation} suppose that the potential
\eqreff{multiv} is actually an approximation to the sum of
sinusoidal potentials, evaluated near the minimum of the
potential. They called this
 $N$-flation.
 Such a potential might arise in string theory, as the sum of
 the potentials of axions which are  components of complex moduli, and
it is argued  that they are under good theoretical control.

Using the $\delta N$ results, the  prediction for  \eq{multiv} is very simple
\cite{treview}. One finds
\be
N(k,\phi_i,\rho)= \frac14 \sum\phi_i^2(k) \label{Nmulti}
\,,
\ee
independently of the final epoch if it is  well after horizon exit.\footnote
{In view  of this independence, one does not expect that a significant
contribution to the curvature perturbation will be generated at the
end of inflation, provided that the end corresponds to a failure of
slow roll. The
possible generation of a contribution to the curvature perturbation
at the end of inflation is discussed in \cite{lev,my05}.}
(Remember that $k$ denotes the epoch of horizon exit $aH=k$.)
The predictions are
\bea
\calp_\zeta &=& 2N \(\frac{H}{2\pi} \)^2  \\
r &=& 8/N \\
\epsilon &=& 2\frac
{\sum m_i^4 \phi_i^2}
{ \(\sum m_i^2\phi_i^2 \)^2 } \\
n-1 &=& - 2\epsilon - \frac1 N
\,.
\eea
The predictions for $\calp_\zeta$ and
$r$ are the same as in the single-component case.
The minimum value of $\epsilon$ at fixed $N$ and $\calp_\zeta$
is $1/2N$ and corresponds
to equal masses. This reproduces the single-component result
$n-1=-2/N$. Making the masses unequal increases
$\epsilon$ and one cannot calculate an upper bound on $\epsilon$ without
knowing the regime of parameter space within which continuous slow-roll takes
place. Hence, making the masses unequal decreases  $n-1$ by an unknown amount
while leaving $r$ the same. As seen in Figures \ref{rvsn} and
 \ref{newrvsn},
this decreases the already-marginal viability of the model
when compared with the data that we are using.

Using \eq{Nmulti} one finds that
the  non-gaussianity parameters $\fnl$ and $\tnl$
given by \eqs{fnl}{tnl} are both much less than 1. This  means that $\fnl$
(and presumably also $\tnl$) will never be measurable.\footnote
{After the first version of this paper was released, the prediction
for two-component chaotic inflation was calculated \cite{rst2} using the
formalism of \cite{rst}.
With with $\phi_1
=\phi_2$ and $m_2/m_1=9$, these authors report that $n=0.93$ and
$f\sub{NL}=1.8$. In contrast we find for these parameters $n=0.95$
and (as just stated) $|\fnl|\ll 1$. The origin of this
 discrepancy is not clear.}

The non-gaussianity
parameters have so far been evaluated for just two other multi-component
inflation models, those of  \cite{ks,ev}. Assuming that $\zeta$ attains the
observed value by the end of inflation the non-gaussianity is again found
to be negligible \cite{lr} in the case of \cite{ks},
but it could be significant in the case of
\cite{ev} if the non-gaussian part of the $\zeta$ is generated  only after
inflation is over \cite{lr}.

            \section{Outlook}

\label{s:outlook}

With present data the bounds that we have presented are not very restrictive.
One sees from \eqst{b1}{b3} that it would be quite
reasonable to shift  the allowed interval for $n$ down by $0.02$ or so, which
would weaken the constraints on models presented in Figure \ref{nvsN}.
A downward shift in $n$ would also improve the viability of the large-field
models presented in \eq{rvsn}.

The situation will change dramatically
when the accuracy of the data is improved.
Consider first the possibility of discriminating among small-field models,
through their prediction for the spectral tilt.
Data from PLANCK \cite{planck}
should give  \cite{kks} $\Delta n= \pm 0.007$, reducing to $\Delta n =.003$
with the  proposed CMBpol \cite{cmbpol}. Adding galaxy survey data should
give further reduction.
Looking at Figure \ref{nvsN}, one
 can distinguish  four  possibilities, according to the eventual value
of $n$.

\paragraph{A value of $n\lsim 0.98$}
 This case is very interesting because it is the expected
prediction for some of the
most reasonable-looking potentials.  These are the quadratic potential
\eqreff{vquad}, the
potential \eqreff{vnew} with $p\geq 2$ (corresponding to  new
and modular inflation) the same potential with $p<0$ (corresponding to mutated
hybrid inflation, some $D$-brane proposals and an $N=2$ supergravity proposal),
the $D$/$F$-term potential \eqreff{loop} and the exponential potential
\eqreff{vqexp}. Regarding the last two  as special cases corresponding to
$p\to 0$ and $p\to \infty$ we are dealing with the potential
\be
V = V_0 \[ 1- \(\frac\phi\mu\)^p \]
\label{vnew2}
\,,
\ee
with $\mu\lsim \mpl$ and $V_0$ dominating.

For $p=2$, the spectral tilt has the scale-invariant value $n=-
2V_0/\mpl^2\mu^2$ provided that this potential is valid on cosmological
scales.  For all of the other relevant values of $p$ ($p\gsim 3$ and $p\leq 0$)
we suppose that the potential is valid also
for long enough that  it gives \eq{Nnew} for $N$.
Then  the spectral tilt depends only on $p$, being given by
 \eq{nnew}.
 Adopting the reasonable range \eq{Nest3}, and taking
$V\quarter\simeq10^{16}\GeV$ which for this type of potential
is more or less required by the  normalization of the spectrum, this
 prediction  becomes
\be
n-1 = -( 0.037 \pm 0.005) \frac{p-1}{p-2}
\,.
\ee
For the allowed range of $p$ the fraction is bigger than $1/2$, and observation
requires it to be $\lsim 2$. We conclude that {\em the theoretical uncertainty
in this case is roughly $\Delta n= 0.005$, similar to the accuracy promised by
PLANCK data.}

%As seen in  Figure \ref{pvsn} this prediction combined with the accuracy
% promised by PLANCK data will
%give a useful  measurement of the shape parameter $p$.
%Any  subsequent improvement in the  accuracy of
%$n$  will not much improve the measurement of $p$ unless  the  theoretical
%uncertainty can be reduced. This uncertainty comes from the uncertainty
%in $N$, and to reduce it one will need a  model of the early
%Universe after inflation. The model would in turn be subject to other
%constraints, coming from a range of astronomical, collider and detector
%constraints. In that situation $n$ would be one of several parameters, the
%whole set to be constrained using all relevant data.

Since the prediction for $n-1$ is proportional to $1/N$, the predicted running
$n'\equiv dn/d\ln k$ is
\be
n' = (n-1)/N  = -   \frac {p-1}{p-2} \frac 2{N^2} ,
\label{nprimenew}
\,,
\ee
which will be in the range $-.0004$ to $-.0016$ or so.
The expected accuracy \cite{kks} is
$\Delta n' \sim 0.003$ from PLANCK
and  $\Delta n' = .0017$ from CMBpol, and galaxy survey
data will reduce these uncertainties. It is therefore possible that the
predicted running  can be verified.

\paragraph{A  value $0.98\lsim n<1.00$.}
This   will indicate a concave-downward
potential. If we are dealing with a  non-hybrid
model the potential cannot have
 the simple parameterization \eqreff{vnew} ($p>2$ or $p<0$).
% That would be possible though if the potential refers to a hybrid model.

\paragraph{A value of $n$ indistinguishable from 1}
This would be compatible both with concave-downward hybrid inflation
potentials, and with the original
 hybrid inflation potential
 $V=V_0+ \frac12 m^2\phi^2$. It would  also be compatible with some more
complicated models \cite{steve,ks}, and the poorly-motivated
concave-upward potentials
of the form \eqreff{vstrange}.

\paragraph{A value $n>1$} This would favour the original hybrid inflation model
$V=V_0+\frac12m^2\phi^2$ or its running-mass version \cite{ewanrunning}.
The  concave-upward potential \eqreff{vstrange} could also give
$n$ significantly above 1, but not for the most reasonable case
 $N\sub{tot}>> N$. A measurement of the running might provide discrimination
between these possibilities.

Now consider the  tensor fraction $r$.
 PLANCK \cite{planck} will give $r\lsim 0.05$.
        Clover \cite{clover}
will give something like $r<10^{-2}$ and the eventual
        limit may  be \cite{rlimit}
something like $r<10^{-4}$. We see from Figure \ref{newrvsn} that
the tensor fraction generated by  small-field models is very small.
It  will not be
detectable by PLANCK or Clover, and  it will never be detectable
if the small-field condition is well-satisfied as would be desirable in
the context of the effective field theory framework described in Section
\ref{effective}.

If   $r$ is indeed   observed by PLANCK
or Clover,  that   would definitely require a large-field model. At present the
only large-field models that are under serious
consideration and are compatible with
observation are Natural Inflation (corresponding to
 a sinusoidal potential),   its limit
single-component Chaotic Inflation
(corresponding to $V\propto \phi^2$) and multi-component Chaotic Inflation.

Natural Inflation corresponds to a line in the $r$-$n$ plane, with
an  endpoint corresponding  to single-component Chaotic Inflation.
The same statement applies to
multi-component Chaotic Inflation. Single-component Chaotic Inflation
corresponds to the point $n= 1 - 2/N$ and $r = 8/N$, with
 $V\quarter$ not far below $10^{16}\GeV$. Using the reasonable
range \eq{Nest3} the predictions become
\be
n= - 0.963 \pm 0.005 \qquad r = 0.14 \pm 0.02
\,.
\ee
Both  $n-1$ and $r$ are  proportional to $1/N$,
 their fractional
uncertainties are given by \eq{delNoverN} as roughly $10\%$. Also,
their scale dependence is $\propto 1/\ln k$ which might eventually
be detectable. To agree with observation the parameters of Natural
Inflation and multi-component Chaotic Inflation should be chosen
so that
 the prediction is not vastly different from this one.

Finally, it is worth saying again that any detection of the non-gaussianity
parameter $f\sub{NL}$ would rule out all single-component slow-roll inflation
models, as well as multi-component Chaotic Inflation, on the hypothesis that
 the inflaton perturbation  generates the curvature perturbation.

\section{Conclusion}
\label{conclusion}

In this paper we have surveyed most forms of  the inflationary potential,
which have some motivation in the context of current ideas about field theory
 beyond the Standard Model of particle physics. Assuming that the inflaton
generates the curvature perturbation,
 we have seen how observation
can constrain the parameters for a given form
of the potential, or rule it out.
% The projected accuracy of measurements of the
%spectral tilt  and the tensor fraction will allow this approach to be
%essentially completed within a few years (apart from the creation of new
%models) since beyond that point
%the accuracy of the observations will go below the
%uncertainty of the predictions. Further advance will come only by confronting
%a more complete model of the early Universe with observation, providing
%a constraint on $N$.

The approach we are taking  may be contrasted with the two other
 lines of inquiry.
One of them \cite{reconstruct}
asks to what extent one may reconstruct the inflationary
potential given an essentially unlimited amount of information about the
spectrum $\calp_\zeta(k)$,  and preferably also about the tensor fraction
$r(k)$. The other \cite{flow} generates, using `flow equations' a large
family of potentials which can satisfy the data, without reference to
current ideas about  field theory.

            \begin{acknowledgments}
            DHL thanks      Andrew Liddle for useful correspondence about the
            prospects for measuring the tensor fraction, and
we thank Andrei Linde
for useful comments on an earlier draft. DHL
is supported by
     PPARC grants PPA/G/S/2002/00098, PPA/G/O/2002/00469, PPA/Y/S/2002/00272,
            PPA/V/S/2003/00104  and EU grant MRTN-CT-2004-503369.
            LA thanks Lancaster University for the award of the studentship from Dowager Countess Eleanor Peel
            Trust.
\end{acknowledgments}

\end{document}